\documentclass[aps,prx,superscriptaddress,twocolumn,preprintnumbers,amsmath,amssymb]{revtex4-2}
\usepackage{color}
\usepackage{array}
\usepackage{amsmath}
\usepackage{amssymb}
\usepackage{graphicx}
\usepackage{pifont}
\usepackage{epstopdf}
\usepackage{bbm}
\usepackage{hyperref}
\usepackage{float}
\usepackage{bibentry}
\usepackage{tcolorbox}
\usepackage{bm}
\usepackage{dsfont}
\usepackage{mathrsfs}
\usepackage{multirow}

\newcommand{\cmark}{\ding{51}} 
\newcommand{\xmark}{\ding{55}} 

\usepackage{braket}
\usepackage{simpler-wick}

\begin{document}
\title{Interaction induced flattening of optical transition quantum geometry}
\author{Xu Yang}
\email{xu.yang@ntu.edu.sg}
\affiliation{Division of Physics and Applied Physics, School of Physical and Mathematical Sciences, Nanyang Technological University, Singapore 637371}
\author{Justin C. W. Song}
\email{justinsong@ntu.edu.sg}
\affiliation{Division of Physics and Applied Physics, School of Physical and Mathematical Sciences, Nanyang Technological University, Singapore 637371}
\date{\today}
\begin{abstract}
The Riemannian geometry of optical transition dipoles has become a useful picture for understanding linear and nonlinear optical response. Here we argue that the interacting quantum geometry of optical transitions possesses a rich structure and can be naturally delineated into two types: localized and delocalized particle-hole excitations. The former possess uniform quantum geometry with flat (vanishing) Hermitian curvature; the latter possess non-uniform quantum geometry with non-vanishing Hermitian curvature. As a striking example, we find that uniform quantum geometry can be produced by electron-hole interactions: even when composed from extended Bloch states in the particle and hole bands, we find excitons have uniform and flat quantum geometry. By developing a many-body length gauge formulation of nonlinear response, we find this uniform and flat excitonic quantum geometry zeros its third-order circular photoconductivity in non-magnetic materials in stark contrast to its non-interacting counterparts. Similarly, the zero Hermitian curvature of localized optical transitions locks their Hall response to that of the ground state. This demonstrates the rich landscape of many-body optical response controlled by an interacting quantum geometry.
\end{abstract}
\maketitle

\textcolor{blue}{{\it Introduction.}}
The resonant quantum geometry of optical transitions has become a succinct framework for describing optical phenomena in terms of the geometrical characteristics of pairs of Bloch states \cite{ahn2022riemannian}. For example, the rate of interband optical absorption is captured by the optical transition Hermitian metric\cite{souza2000polarization,ahn2022riemannian}, the second-order shift photocurrent \cite{sipe2000second,ahn2020low,ahn2022riemannian,ma2023photocurrent} by the Hermitian connection, and the third-order photoconductivity can be dominated by its Hermitian curvature\cite{ahn2022riemannian}. However, the study of resonant quantum geometry has, so far, focused on non-interacting electrons. What is the effect of electron interactions in determining resonant quantum geometry?

At the core of this important question is the systematic connection between resonant quantum geometry and optical response\cite{morimoto2016topological,nagaosa2017concept,ahn2022riemannian,mao2025low}. For non-interacting electrons, this connecting bridge was established via the length-gauge (position operator) framework~\cite{blount1962formalisms,aversa1995nonlinear}. Directly extending this formulation to interacting many-body systems, however, presents conceptual challenges~\cite{resta1998quantum,resta2024}. Instead, optical response of interacting many-body electronic materials are more naturally formulated in terms of the velocity gauge \cite{kubo1957statistical,passos2018nonlinear,rostami2021gauge} that are plagued with unphysical divergences for higher order nonlinear susceptibilities\cite{aversa1995nonlinear}. As a result, correspondence between nonlinear optical response and interacting geometrical characteristics have only been established in isolated cases \cite{resta2022theory,resta2024,matsyshyn2026superconducting,guan2026exploring,yang2026correlated}.

Here we find that electron interactions can qualitatively reshape the resonant quantum geometry of optical transitions producing striking departures from that of a non-interacting description of optical transitions. In so doing, we develop a many-body length gauge perturbation theory for nonlinear optical response enabling a systematic correspondence of nonlinear susceptibilities with many-body resonant quantum geometrical characteristics. 
Specifically, we find that interactions naturally produce two types of resonant quantum geometry: uniform quantum geometry and non-uniform quantum geometry describing localized and delocalized particle-hole excitations respectively. The former are characterized by a uniform Hermitian metric and Hermitian connection as well as a vanishing Hermitian curvature (i.e. flat) even when their underlying non-interacting Bloch states are extended. The latter non-uniform quantum geometry possess a non-trivial holonomy.

The interaction induced flat and uniform quantum geometry are naturally realized in bound excitons (localized particle-hole excitations) below the many-body particle-hole continuum\cite{wannier1937structure}. Indeed, as a concrete application of our many-body length gauge framework, we find that the exciton transition's flat and uniform quantum geometry zeroes its (third-order nonlinear) circular photoconductivity in non-magnetic materials, e.g., in transition metal dichalcogenides (TMDs)\cite{wang2018colloquium}. This dramatically constrains its role in producing Hall photoconductivity e.g., in experiments of valley selective Hall photoconductivity in TMDs~\cite{mak2014valley,ubrig2017microscopic}. This strongly contrasts with that of free particle-hole excitations which possess large circular photoconductivity that dominate the Hall photoconductive response of non-magnetic materials \cite{yao2008valley,song2016giant,yin2022tunable,ahn2022riemannian}. Together this demonstrates the rich variety of nonlinear optical response that can be obtained with many-body resonant quantum geometry.

\textcolor{blue}{{\it Many-body resonant quantum geometry and hermitian curvature.}} Resonant optical transitions are a central process of optical response \cite{haug1994quantum}. Unlike non-interacting electrons, where transition dipoles and its associated quantum geometry are well described by single-particle electron momentum~\cite{ahn2022riemannian}, crystal momentum $\textbf{k}$ of an individual particle is no longer generically conserved in interacting electronic systems; electron-electron interactions enables mixing between states with different $\textbf{k}$. Instead, the natural quantity parameterizing the transition dipole in many-body systems is flux $\bm{\kappa}$~\cite{kohn1964theory,niu1985quantized,souza2000polarization,watanabe2018insensitivity,resta2024}. Here $\bm\kappa$ is a static uniform vector potential across the entire system that enables to parameterize the many-body Hamiltonian $\hat{H}(\bm{\kappa})$. In this way, many-body eigenstates $\ket{\Phi_n (\bm{\kappa})}$ and eigen-energies $E_n (\bm{\kappa})$ can then be labeled by $\bm{\kappa}$~\cite{kohn1964theory}. 

Importantly, the transition dipole between many-body eigenstates can also be parameterized by $\bm \kappa$. Writing $\hat{\bm v} = i[\hat{H},\hat{\textbf{r}}]/\hbar$ and $\hat{\bm v} = \hbar^{-1}\partial_{\bm \kappa} \hat{H}$, the many-body transition dipole matrix elements between the ground state $\ket{\Phi_0 (\bm\kappa)}$ and an excited state $\ket{\Phi_n (\bm\kappa)}$ can be written as~\cite{yang2026correlated} 
\begin{equation}
    \bm{r}_{n0} (\bm \kappa)=  \hspace{-0.5mm}\bra{\Phi_n (\bm\kappa)}i\partial_{\bm\kappa} \ket{\Phi_0 (\bm\kappa)}, \,\, \hat{e}_a^{n0}(\bm{\kappa})= P_n(\bm{\kappa})i\partial_{a}P_0(\bm{\kappa})
\end{equation}
where $\hat{e}_a^{n0}(\bm{\kappa})$ is a many-body tangent vector that enables a compact description of optical transitions and we use the shorthand $\partial_{\kappa_a}\equiv\partial_a$; notice that $\hat{e}_a^{n0}(\bm{\kappa}) = r_{n0} (\bm \kappa) \ket{\Phi_n (\bm\kappa)}\bra{\Phi_0 (\bm\kappa)}$ with $P_{0/n} (\bm \kappa)=\ket{\Phi_{0/n}(\bm \kappa)}\bra{\Phi_{0/n}(\bm \kappa)}$ are projectors formed from the many-body eigenstates. For a detailed explanation in terms of the many-body quantum Liouville equation, see also Supplementary Information {\bf SI}. The tangent vector $\hat{e}_a^{n0}(\bm{\kappa})$ directly tracks the optical transition rate. For example, the rate of optical transitions is $W_{0\to n} =(2\pi e^2/\hbar^2) {\rm lim}_{\bm \kappa \to 0} \delta(\omega_{n0} (\bm \kappa)-\omega) 
Q^{n0}_{ab} (\bm \kappa) E_a(\omega)E_b(-\omega)$, where $Q^{n0}_{ab} (\bm \kappa)\equiv(\hat{e}^{n0}_a (\bm \kappa),\hat{e}^{n0}_b (\bm \kappa))$ is the Hilbert-Schmidt inner product of the tangent vectors
\footnote{The Hilbert-Schmidt inner product is defined as $(A,B)=\text{Tr}(A^{\dagger}B)$.}, $E_{a,b}$ are electric fields, and $\omega_{n0}(\bm \kappa)=\hbar^{-1}(E_n (\bm \kappa)-E_0(\bm \kappa))$; here and below, repeated indices are summed over. Indeed, the resonant Hermitian metric $Q(\bm \kappa)$ closely tracks the real part of the linear in $E$ optical conductivity~\cite{ahn2022riemannian}.

Importantly, the tangent vector can change as $\bm{\kappa}$ is varied: this variation with $\bm \kappa$ defines a many-body resonant quantum geometry. Successive higher order derivatives of the tangent vector track higher order nonlinear optical responses\cite{ahn2022riemannian,Mitscherling2025,avdoshkin2025,yang2026correlated}. For example, as the tangent vector is parallel transported in $\bm \kappa$, its direction and magnitude can change (see Figure 1a) and can be tracked by $\nabla_{\bm{\kappa}}\hat{e}_{a}^{n0}(\bm{\kappa}) = P_n(\bm{\kappa}) (\partial_{\bm{\kappa}} \hat{e}_{a}^{n0}(\bm{\kappa})) P_0(\bm{\kappa})$. Its components are the Hermitian connection $C^{n0}_{bac}(\bm{\kappa}) \equiv (\hat{e}_b^{n0}(\bm{\kappa}), \nabla_{a} \hat{e}_c^{n0}(\bm{\kappa}))$ which produce the shift current second-order nonlinear conductivity \cite{ahn2022riemannian,avdoshkin2025,yang2026correlated}.

\begin{figure}
    \centering
    \includegraphics[width=1.0\linewidth]{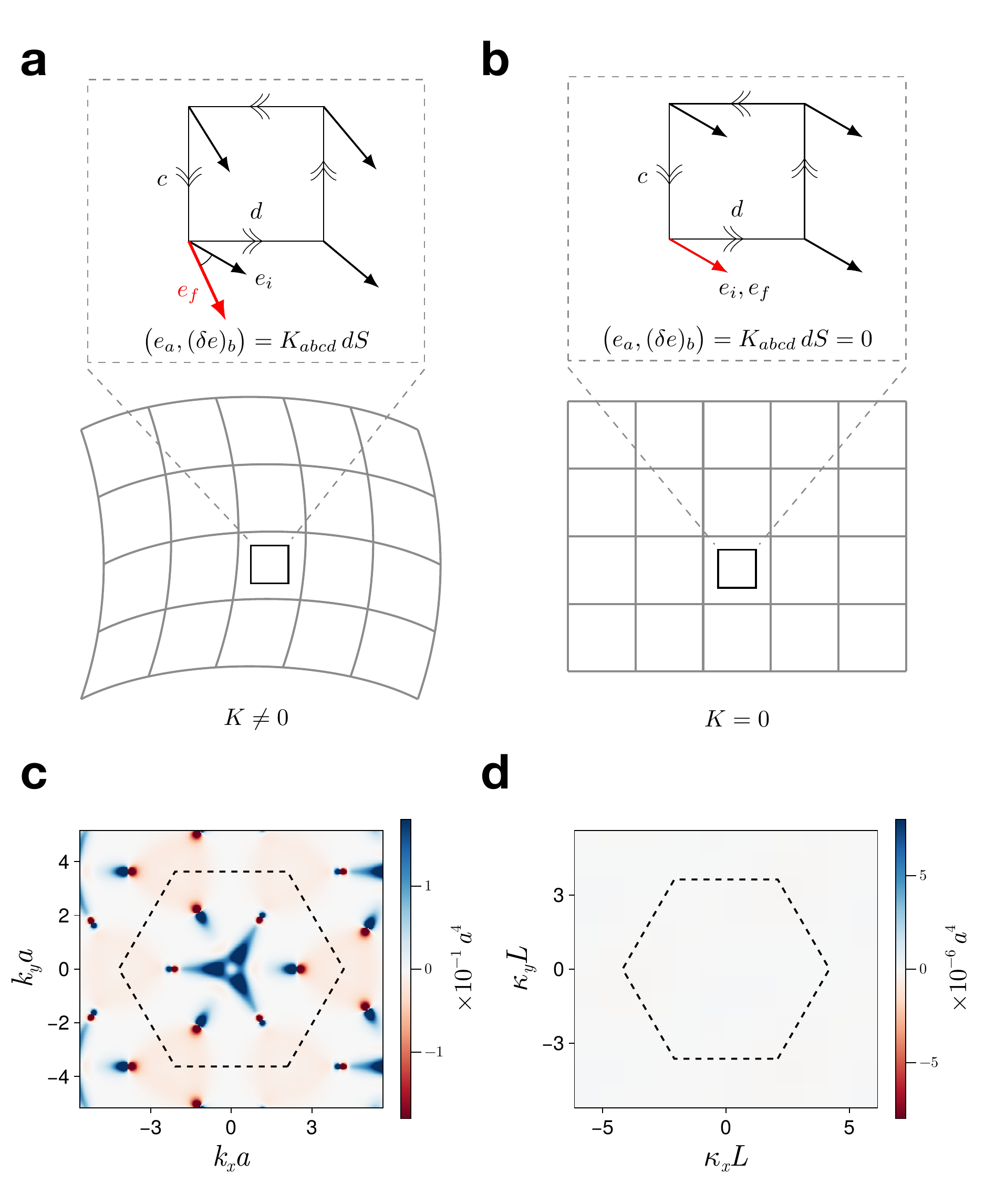}
    \caption{\textcolor{blue}{\it Non-flat vs. interaction induced flat resonant quantum geometry.} {\bf a.} Parallel transport around an infinitesimal loop on a non-flat resonant transition manifold. The mismatch between the initial tangent vector $e_i$ and its parallel-transported counterpart $e_f$ around the square spanned by the $c$ and $d$ directions is encoded in $\delta e=e_f-e_i$ and governed by the Hermitian curvature tensor. {\bf b.} Parallel transport on a flat resonant transition manifold. The tangent vector is unchanged after transport, corresponding to $K=0$. {\bf c.} Delocalized free optical transition Hermitian curvature tensor $K^{cv}_{[x,y]xy}$ (antisymmetrized in the first two indices, relevant to the photovoltaic Hall response) in the Brillouin zone for unstrained three-band MoS$_2$, computed between the valence and lower conduction bands. {\bf d.} Excitonic optical transition Hermitian curvature tensor $K^{n_{\rm ex}0}_{[x,y]xy}$ in flux space for the same system, computed between the ground state and the first exciton for a $48\times 48$ real-space lattice system size.}
    \label{fig1}
\end{figure}

How curved is this manifold of resonant transitions? This is captured by the Hermitian curvature tensor $K^{n0}_{abcd}(\bm{\kappa})$ obtained by tracking how a tangent vector changes across an infinitesimal loop \cite{schutz1980geometrical,nakahara2018geometry} $K^{n0}_{abcd}(\bm{\kappa})=(\hat{e}^{n0}_{a}(\bm{\kappa}),(\nabla_c\nabla_d-\nabla_d\nabla_c)\hat{e}^{n0}_b(\bm{\kappa}))$. Geometrically, it describes the deviation of $\hat{e}_b(\bm{\kappa})$ after being parallel transported around an infinitesimal parallelogram spanned by directions $c$ and $d$ (see Fig.~\ref{fig1}(a)). The Hermitian curvature tensor can be expressed in terms of derivatives of Hermitian connections as:
\begin{align}\label{eq:K_expression}
K^{n0}_{abcd} (\bm \kappa) \equiv\partial_cC^{n0}_{adb} (\bm \kappa) \hspace{-0.5 mm}-\hspace{-0.5 mm}\partial_dC^{n0}_{acb} (\bm \kappa) \hspace{-0.5 mm}-\hspace{-0.5 mm}S^{n0}_{ab[c,d]} (\bm \kappa) \hspace{-0.5 mm} ,
\end{align}
where $S^{n0}_{abcd}(\bm{\kappa})=(\nabla_c\hat{e}^{n0}_a(\bm{\kappa}),\nabla_d\hat{e}^{n0}_b(\bm{\kappa}))$ measures the quadratic change in $Q$ under parallel transport, and $S^{n0}_{ab[c,d]}(\bm{\kappa})=S^{n0}_{abcd}(\bm{\kappa})-S^{n0}_{abdc}(\bm{\kappa})$ is anti-symmetrized over $c,d$. In proving Eq.~\eqref{eq:K_expression}, we use the relation $\partial_{c}(e_a,e_b)=(\nabla_ce_a,e_b)+(e_a,\nabla_ce_b)$ valid for any two tangent vectors $e_a,e_b$. For a flat (curved) geometry $K=0$ ($K\neq 0$), tangent vectors return to themselves (do not return to themselves) after a closed loop yielding a trivial (non-trivial) holonomy\cite{nakahara2018geometry}. At a physical level, as we shall see, the Hermitian curvature tensor controls the photoconductivity, which is a third order nonlinear response.

\textcolor{blue}{{\it Uniform and flat quantum geometry for localized excitation transitions.}}
We now argue that the localized nature of particle-hole pair amplitude of an excitation on resonance strongly constrains the quantum geometry of the optical transition. For an excitation $\ket{\Phi_n}$, we consider its projection onto one-particle-hole pair sector: $\Psi(\bm{x}_e,\bm{x}_h)=\bra{\Phi_0}a_{\bm{x}_h}^{\dagger}a_{\bm{x}_e}\ket{\Phi_n}$, where $a^{\dagger}/a$ is the electron creation/annihilation operator. We call the excitation charge localized if its particle-hole amplitude decays exponentially at large electron-hole separation: $\Psi(\bm{x}_e,\bm{x}_h)\sim e^{-|\bm{x}_e-\bm{x}_h|/\xi}$ known as relative coordinate localization. Critically, this relative coordinate localization produces a flux transformation relation as~\cite{yang2026correlated}: 
\begin{equation}
\Psi^{\rm loc}_{\bm{\kappa}}(\bm{x}_e,\bm{x}_h)=e^{-i\bm{\kappa}\cdot (\bm{x}_e-\bm{x}_h)}\Psi^{\rm loc}_{\bm{0}}(\bm{x}_e,\bm{x}_h) 
\label{eq:local}
\end{equation}
in the thermodynamic limit. Consequently, the excitation energy is also exponentially insensitive to flux: $\partial_{\kappa}E_{n_{\mathrm{loc}}0}=0$. A prime example of such relative localized excitations are excitons (bound electron hole pairs) that have energy below the particle-hole continuum (single-particle band gap)\cite{wannier1937structure}. Notice that the exciton is relative coordinate localized even as its component electron and hole arise from extended Bloch states. Such relative coordinate localization are not confined to excitons alone; they can also be achieved for magnons \cite{olsen2021unified,esquembre2025magnons}, particle-hole excitations between impurity states, and  Anderson localized states. As we now argue, the flux threading properties in Eq.~(\ref{eq:local}) severely constrain the interacting quantum geometry of such optical transitions producing a uniform and flat resonant quantum geometry.

To see this, we first examine the Hermitian metric, which is determined by $\Psi(\bm{x}_e,\bm{x}_h)$ as:
\begin{align}
Q^{n0}_{ab} (\bm \kappa)\!=\!r^a_{0n} (\bm \kappa) r^b_{n0} (\bm \kappa),\!\!\quad \bm{r}_{0n} (\bm \kappa)\!=\! i 
\!\int\! d\bm{x}\,\hat{\bm{v}}_e\,\!\frac{\Psi_{\bm \kappa}(\bm{x},\bm{x})}{\omega_{n0}(\bm \kappa)}\!.
\label{eq:Q}
\end{align}
Here $\hat{\bm{v}}_e$ is the single-particle velocity operators acting on the coordinates $\bm{x}_{e}$. Although $r_{0n}$ involves the full flux dependence of the many-body state $\ket{\Phi_n}$, the optical transition only excites a single particle-hole pair, and therefore we can use $i\omega_{0n}r_{0n}=v_{0n}$ to convert it to a functional form involving only the one-p-h-amplitude. Inserting the transformation rules of $\Psi$ under flux insertion, we find $Q^{n_{\rm loc}0}_{ab}(\bm{\kappa})=Q^{n_{\rm loc}0}_{ab}(\bm{0})$. This follows because $E_{n_{\rm loc}0} (\bm \kappa)$ is flux independent, and the exciton wavefunction phase factor ${\rm exp}[-i\bm{\kappa}\cdot (\bm{x}_e-\bm{x}_h)]$ accumulated under flux insertion cancels when $\bm{x}_e=\bm{x}_h =\bm{x}$. Additionally, noticing the identity $Q^{n0}_{ab}(\bm \kappa)S^{n0}_{ab[c,d]}(\bm \kappa)=(\partial_cQ^{n0}_{ab}(\bm \kappa))C^{n0}_{adb}(\bm \kappa)-(\partial_dQ^{n0}_{ab}(\bm \kappa))C^{n0}_{acb}(\bm \kappa)$, the $\bm \kappa$ insensitivity of the localized transition Hermitian metric $\partial_a Q^{n_{\rm loc}0}_{bc}(\bm \kappa)=0 $ means that $S^{n_{\rm loc}0}_{abcd}(\bm \kappa)=S^{n_{\rm loc}0}_{abdc}(\bm \kappa)$. 

We next examine the flux dependence of the Hermitian connection $C^{n0}_{bcd} (\bm \kappa)$. The explicit form of $C^{n0}_{bcd}(\bm \kappa)$ in terms of transition matrix elements and the excitation wave-function is (we defer detailed derivation to \textbf{SI}):
\begin{align}\label{eq:C_expression}
&C^{n0}_{bcd}(\bm{\kappa})= r^b_{0n}(\bm{\kappa})\partial_c(r^d_{n0}(\bm{\kappa}))-ir^b_{0n}(\bm{\kappa})r^d_{n0}(\bm{\kappa})d^{c}_{0n,\bm{\kappa}},
\end{align}
where $\bm{d}_{0n,\bm{\kappa}} = \mathcal{\bm{A}}_n(\bm{\kappa}) - \mathcal{\bm{A}}_0(\bm{\kappa})$, with $\mathcal{\bm{A}}_{n}(\bm{\kappa})=\bra{\Phi_n(\bm{\kappa})}i\partial_{\bm{\kappa}}\ket{\Phi_n(\bm{\kappa})}$ (and similarly for $\mathcal{\bm{A}}_0$) being the many-body Berry connection. For Slater determinant states, $\bm{d}_{0n}$ takes on a simple form: $i\int d\bm{x}_ed\bm{x}_h \Psi_{\bm{\kappa}}^{*}(\bm{x}_e,\bm{x}_h)\partial_{\bm{\kappa}}\Psi_{\bm{\kappa}}(\bm{x}_e,\bm{x}_h)$. Following the same reasoning as above, and the fact that in the last term the flux-dependent phase acquired by $\Psi^{\rm loc}_{\bm \kappa}$ is exactly canceled by the conjugate phase in $[\Psi_{\bm \kappa}^{\rm loc}]^*$, we find that $C^{n_{\rm loc}0}_{bcd} (\bm \kappa)$ is independent of $\bm{\kappa}$, i.e., $\partial_a [C^{n_{\rm loc}0}_{bcd}(\bm \kappa)]=0$ for all $\bm{\kappa}$. Because $\partial_a [C^{n_{\rm loc}0}_{bcd}(\bm \kappa)]=0$ and $\partial_a Q^{n_{\rm loc}0}_{bc}(\bm \kappa)=0 $, the resonant quantum geometry for such localized optical transitions is {\it uniform}. 

Substituting $S^{n_{\rm loc}0}_{abcd} (\bm \kappa) =S^{n_{\rm loc} 0}_{abdc} (\bm \kappa) $ and the $\bm{\kappa}$ insensitivity of $C^{n_{\rm loc}0}_{bcd}(\bm \kappa)$ into Eq.~\eqref{eq:K_expression}, we obtain two key relations:
\begin{align}\label{eq:K_and_Omega}
K^{n_{\rm loc}0}_{abcd} (\bm \kappa) =0,\quad \Omega^{cd}_{n_{\rm loc}}(\bm{\kappa})=\Omega^{cd}_0(\bm{\kappa}),
\end{align}
where $\Omega^{cd}_{n/0}\equiv  i\text{Tr}[P_{n/0}(\partial_cP_{n/0})(\partial_dP_{n/0})]-(c\leftrightarrow d)$ is the Berry curvature for state $\Phi_{n/0}$. Eq.~\eqref{eq:K_and_Omega} is the central result of our paper. The first relation $K^{n_{\rm loc}0}_{abcd}=0$ implies a {\it flat} resonant quantum geometry for transitions between the ground state and a localized excitation. The second relation can be understood from the general identity $K^{n0}_{abcd}(\bm{\kappa})=-iQ^{n0}_{ab}(\bm{\kappa})(\Omega^{cd}_n(\bm{\kappa})-\Omega^{cd}_0(\bm{\kappa}))$, which connects the ground state quantum geometry and the localized excited state quantum geometry via the resonant quantum geometric object $K$. Consequently, the localized excitation Berry curvature $\Omega_{n_{\rm loc}}$ is locked to that of the ground state $\Omega_0$.

Physically the Hermitian curvature tensor describes the optical pumping of Berry curvature at a rate controlled by $Q$\cite{fregoso2019bulk,fregoso2020erratum,ahn2022riemannian} and therefore contributes to the light-induced DC Hall response, i.e., the photovoltaic Hall effects\cite{song2016giant,mak2014valley,ubrig2017microscopic,yin2022tunable,yao2008valley}. For non-interacting delocalized particle-hole excitations, $K^{n_{\rm free}0}$ is generally finite, 
indicating a non-flat resonant quantum geometry as shown in Fig.~\ref{fig1}(c); its momentum integral over an on-resonant contour does not vanish\cite{sun2025band} yielding a finite photovoltaic Hall response. Such a contribution can be particularly important in massive Dirac systems near the band edge~\cite{ahn2022riemannian}, where resonant conduction- and valence-band electrons carry large Berry curvatures of opposite signs. Eq.~ (\ref{eq:K_and_Omega}) indicates that localized optical transitions (such as bound excitons) behave fundamentally different. Their Hermitian curvature tensor vanishes identically over the entire flux manifold, $K^{n_{\rm loc}0}_{abcd}=0$, thereby strongly constraining the photovoltaic Hall response. As we show below, the implication is in fact stronger: the full (third order) injection Hall photoconductivity vanishes for localized optical transitions such as excitons in stark contrast to free particles. 

The relation $\Omega_{n_{\rm loc}}=\Omega_0$ for localized excitations is also quite striking. To appreciate this consider an exciton that is a coherent superposition of conduction electrons and valence holes that can carry very different Berry curvatures (e.g., in massive Dirac systems they have opposite signs), but Eq.~\eqref{eq:K_and_Omega} shows that the bound nature of excitons effectively screens the net Berry curvature of the electron-hole pair, locking the exciton Berry curvature to that of the ground state. This can be understood from flux insertion: the many-body Berry curvature measures the transverse response of wave-function to flux insertion\cite{niu1985quantized,watanabe2018insensitivity}, while a bound exciton localized in its relative coordinate is insensitive to flux insertions in the thermodynamic limit. Consequently, it contributes no additional Berry curvature relative to the ground state, yielding $\Omega_{n_{\rm loc}}=\Omega_0$. This is reminiscent of the integer quantum Hall effects, where sweeping the Fermi energy through localized states doesn't change the Chern number, which changes only upon crossing delocalized states in the Landau bands\cite{laughlin1981quantized,halperin1982quantized,bellissard1994noncommutative}.

For a time-reversal-invariant ground state, this implies that a bound exciton contributes no additional Berry curvature and cannot directly generate injection Hall photoconductivity. Consistent with this picture, experimentally observed valley-selective Hall photoconductivity in TMDs has been attributed to charged trions or to the dissociation of excitons\cite{mak2014valley,ubrig2017microscopic}.

\begin{table}[t]
\centering
\begin{tabular*}{\linewidth}{@{\extracolsep{\fill}} c c c}
\hline\hline
Quantity
& Localized
& Delocalized \\ 
\hline
$Q^{n0}_{ab}(\bm{\kappa})$ 
& $\bm{\kappa}$-independent
& $\bm{\kappa}$-dependent \\[5pt]
$C^{n0}_{abc}(\bm{\kappa})$ 
& $\bm{\kappa}$-independent
& $\bm{\kappa}$-dependent \\[5pt]
$S^{n0}_{abcd}(\bm{\kappa})$ 
& symmetric in $c\leftrightarrow d$
& no constraints \\[5pt]
$K^{n0}_{abcd}(\bm{\kappa})$ 
& vanishes
& does not vanish \\[5pt]
\hline\hline
\end{tabular*}
\caption{Constraints for interacting resonant quantum geometry for localized optical transitions (such as excitonic transitions) and delocalized optical transitions. The middle column tabulates the severe constraints that interactions induce.} 
\label{tab:resonant_qg}
\end{table}

\textcolor{blue}{{\it Many-body length-gauge perturbation theory.}}
We now connect resonant quantum geometry to nonlinear optical responses by developing a many-body length-gauge perturbation theory. As we will see below, this allows to directly connect nonlinear optical response with resonant interacting quantum geometry. We adopt a constructive approach by starting from the velocity-gauge perturbation theory and deriving the length-gauge formula for an insulator. In the velocity gauge, the Hamiltonian $H(\bm{A})$ is minimally coupled to a time-dependent vector potential $\bm{A}(t)$ and the density matrix $\rho^A$ then obeys the Liouville equation: $i\hbar \text{d}_t\rho^A=[H(\bm{A}),\rho^{A}]$. Although this formulation can be applied to general many-body systems, physical observables are expressed in terms of the velocity operators $\bm{v}$, which takes a complicated form obscuring geometrical meaning.

We therefore instead treat $\bm{A}$ as a time-dependent parameter of the Hamiltonian, and express the density matrix dynamics through the $\bm{A}$-dependence of the eigenstates and eigen-energies. This is achieved using the instantaneous eigen-basis: $H(\bm{A})\ket{\Phi_n(\bm{A})}=E_n(\bm{A})\ket{\Phi_n(\bm{A})}$, in which we define $\rho^E_{nm,\bm A}=\bra{\Phi_n(\bm A)}\rho^A\ket{\Phi_m(\bm A)}$~\cite{sipe1993nonlinear,ventura2017gauge}. Since $\bm{A}$ itself is time-dependent, we can separate the time-dependence on $\bm E$ from the adiabatic dependence through $\bm{A}$ using $\text{d}_t=\partial_t|_{\bm{A}}+\frac{e}{\hbar}\dot{\bm{A}}\cdot\partial_{\bm{\kappa}}$, where $\partial_{\bm{\kappa}}$ denotes differentiating with adiabatic flux parameter. Introducing the $\bm{A}$-dependent Berry connection $\bm{\xi}_{nm}(\bm A)=\bra{\Phi_n(\bm A)}i\partial_{\bm{\kappa}}\ket{\Phi_m(\bm A)}$ to account for the basis change, the resulting equation of motion can be solved recursively in powers of $\bm{E}$. Starting from the initial condition $\rho^{E,(0)}_{nm}(\bm{A})=\delta_{n0}\delta_{m0}$, the $N$-th order density matrix is obtained from the $(N-1)$th order one as:
\begin{align}\label{eq:rho^E_recursive}
&\rho^{E,(N)}_{nm,\bm{A}}(\omega)\!=\!\frac{ie\bm{E}(\omega_1)\cdot \big[\bm{D}_{\bm A}\rho^{E,(N-1)}_{\bm A}(\omega-\omega_1)\big]_{nm}}{\hbar(\omega-\omega_{nm}(\bm{A}))},
\end{align}
where $\big[D^{\alpha}_{\bm{A}}X\big]_{nm}=\partial_{{\kappa}_\alpha}X_{nm}-i[\xi^{\alpha}(\bm{A}),X]_{nm}$. Expressing physical observables in the instantaneous eigenbasis gives: $\langle O\rangle=\text{Tr}[O(\bm A)\rho^A]= \text{Tr}[O^E(\bm A)\rho^E(\bm A)]$. 

Adiabatic flux threading allows the flux-dependent eigenstates to be traced back to the zero-flux eigenstates. After threading an integer multiple of flux quantum, the Hamiltonian is related to the flux-free Hamiltonian by a gauge transformation and hence has the same energy spectrum~\cite{kohn1964theory}. The eigenstates themselves, however, need not return to the same labels and may undergo spectral flow. We encode this reshuffling by a stitch matrix: $\Pi$: $\Pi_{nm}=\delta_{\pi(n),m}$, where $\pi$ denotes the permutation of eigenstate labels under spectral flow. For an insulator, the ground state is preserved under spectral flow so that $\pi(0)=0$, and the zeroth-order initial condition $\rho^{E,(0)}=\delta_{n0}\delta_{m0}$ satisfies $\rho_{nm}^{E,(0)}(\bm A)=(\Pi\rho^{E,(0)}\Pi^{-1})_{nm}$. Since Eq.~\ref{eq:rho^E_recursive} transforms covariantly under the same relabelling, this relation propagates to arbitrary order of $\rho$, leading to $\rho^{E}(\bm A)=\Pi\rho^{E}\Pi^{-1}$. Similarly, $O^E(\bm{A})=\Pi O^E\Pi^{-1}$.

We can therefore connect the length-gauge expression at $\bm{A}$ to that at no flux: $\text{Tr}(O^E(\bm{A})\rho^E(\bm A))=\text{Tr}(\Pi O^E\Pi^{-1}\Pi\rho^E\Pi^{-1})=\text{Tr}(O^E\rho^E)$ where $\rho^E$ is iteratively solved from the length-gauge equation:
\begin{align}\label{eq:length_gauge}
[i\partial_t-\omega_{nm}] \rho^E_{nm}=i\frac{e}{\hbar}\bm{E}\cdot(\partial_{\bm{\kappa}}\rho^E_{nm}-i[\bm{\xi},\rho^E]_{nm}),
\end{align}
with $\omega_{nm}$ and $\xi_{nm}$ evaluated at $\bm{A}=0$. Eq.~\ref{eq:length_gauge} is the Liouville equation in the many-body length-gauge which is fully determined by the energy and transition dipole matrix element and their flux dependence. Notice that $\bm{\xi}_{n0} (\bm \kappa) = \bm{r}_{n0} (\bm \kappa)$ allowing to directly connect with interacting quantum geometry. A detailed discussion of flux threading can be found in \textbf{SI}. 

The linear response and second-order DC optical responses, including shift and injection currents, follow straightforwardly from this formulation and reproduce results obtained using different formulations~\cite{resta2024,onishi2025quantum,yang2026correlated,guan2026exploring}. We next turn to the third-order optical response.

\textcolor{blue}{{\it Third-order jerk, injection and shift currents.}} We now turn to the DC photoconductivity response (a third-order rectified current response): 
$j^a=
\sigma^{abcd}(0,\omega,-\omega)E^b(0)E^c(\omega)E^d(-\omega)$. According to their distinct relaxation-time $\tau$ dependence, the third-order photoconductivity can be partitioned into jerk ($\propto \tau^2$), injection ($\propto \tau$), and shift ($\propto \tau^0$) responses: $\sigma=\sigma_{\text{jerk}}+\sigma_{\text{inj}}+\sigma_{\text{shift}}$~\cite{fregoso2019bulk,ahn2022riemannian}.

Using the many-body formulation described above and $\bm{j}=-e\text{Tr}(\rho\bm{v})$, we find many-body expressions of each of the third order photoconductivity contributions above; a detailed derivation can be found in {\bf SI}. We first focus on the jerk and injection contributions. We find the many-body third order (photoconductivity) jerk response tensor is given by: $\sigma^{abcd}_{\text{jerk}}= [2\pi e^4\tau^2/(\hbar^3V)]\sum_{n} \delta(\omega-\omega_{n0})\partial_{a}\partial_b(\omega_{n0})Q^{n0}_{cd}$ where $Q^{n0}_{cd}$ is the many-body Hermitian metric (evaluated at $\bm \kappa = 0$)\cite{fregoso2018jerk,ventura2021comment}. 
In the same fashion, we evaluate the many-body third-order injection current contribution to the photoconductivity as: 
\begin{align}
&\sigma^{abcd}_{\rm inj}=\frac{\pi e^4\tau}{\hbar^3V}\sum_n\delta(\omega-\omega_{n0})\Big[iK^{0n}_{dcab}+i\text{d}C\nonumber\\
&+i S^{n0}_{cd[a,b]}-2\partial_a(\omega_{n0})\partial_{E_b}Q^{n0}_{cd}+2i\frac{\partial_a\omega_{n0}}{\omega_{n0}}\mathcal{M}_{cdb}\Big].
\label{eq:injectionmain}
\end{align} 
It is composed of several distinct terms and they are given as follows: the differential term is defined as $\text{d}C=\frac{1}{2}\big[\partial_a(C^{n0}_{cbd}-C^{0n}_{dbc})+(a\leftrightarrow b)\big]$; the field derivative of the Hermitian metric, $\partial_{E_b}Q^{n0}_{cd}$ is defined by perturbing each state using e.g., $\partial_{E_b}\ket{0}\equiv \sum_m \ket{m}{\bra{m}r^b\ket{0}}/{\omega_{0m}}$, which characterizes the change of the Hermitian metric $Q$ under a static electric field $E_b$; and finally the metric–connection term $\mathcal{M}_{cdb}$ is defined as $\big(C^{n0}_{cdb}+(\partial_c\omega_{n0}/{\omega_{n0}})Q^{0n}_{db}\big)-(c\leftrightarrow d)^*$.

The many-body expressions for the jerk and injection current contributions to the $\sigma^{abcd}$ enable straightforward evaluation of responses for localized optical transitions (i.e. $n=n_{\rm loc}$). First, since $\partial_{\bm{\kappa}}(\omega_{n_{\rm loc}0})=0$, the third order jerk current for localized optical transitions vanishes identically. Moving to the injection contribution, we note the same condition also removes the contributions of the field-derivative term $\partial_EQ$ and the metric-connection term $\mathcal{M}$ to the injection current, as both are multiplied by the velocity prefactor $\partial_a \omega_{n_{\rm loc}0}$. 
Similarly, using Table.~\ref{tab:resonant_qg}, we find all the remaining terms in Eq.~(\ref{eq:injectionmain}) also vanish due to the flatness and uniform resonant quantum geometry of localized optical transitions. This means that $\sigma_{\rm inj}^{abcd}$ in Eq.~(\ref{eq:injectionmain}) {\it vanishes} identically for localized optical transitions such as excitonic transitions. This is surprising since the injection current contribution to the photoconductivity is thought to dominate the non-interacting Hall photoconductive response \cite{ahn2022riemannian}.

Finally, we turn to the shift current contribution to the third order photoconductive response which can be obtained in the same way. We defer the general expression to \textbf{SI}. Utilizing the uniformity and flatness of the resonant quantum geometry (Table.~\ref{tab:resonant_qg}), the case for exciton optical transitions can be greatly simplified yielding 
\begin{align}
&\sigma^{abcd}_{{\rm shift,}0\to n_{\rm ex}}=\frac{\pi e^4}{\hbar^3 V}\sum\limits_{n_{\rm ex}}\Bigg[\delta(\omega-\omega_{n_{\rm ex}0})\big[\mathcal{V}-2\mathcal{R}^a_{n_{\rm ex}0} \partial_{E_b}Q^{n_{\rm ex}0}_{cd}\nonumber\\
&+\frac{2}{\omega_{n_{\rm ex}0}}\big(S^{n_{\rm ex}0}_{cbad}+S^{n_{\rm ex}0}_{bdac}\big)\big]+2\partial_{\omega}\delta(\omega-\omega_{n_{\rm ex}0})S^{n_{\rm ex}0}_{cdab}\Bigg],\label{eq:shift_exciton_formula}
\end{align}
where $\mathcal{R}^a_{n_{\rm ex}0}=-iC^{n_{\rm ex}0}_{bab}/Q^{n_{\rm ex}0}_{bb}$ is the excitonic transition shift vector ~\cite{yang2026correlated}, and virtual-transitions are captured by $\mathcal{V}=\sum_{m\neq 0,n_{\rm ex}}\left[r^d_{n_{\rm ex}0}X^{bca}_{0n_{\rm ex},m}+(c\leftrightarrow d)^*\right]$, with $X^{bca}_{0n_{\rm ex},m}\equiv
-iv^a_m[(r^c_{0m}r^b_{mn_{\rm ex}})/\omega_{mn_{\rm ex}}^2+(r^b_{0m}r^c_{mn_{\rm ex}})/{\omega_{0m}^2}]$.

\begin{table}[t]
\small
\setlength{\tabcolsep}{2.5pt}
\begin{tabular}{llcccccc}
\hline\hline
& & \multicolumn{2}{c}{\textbf{Jerk}} & \multicolumn{2}{c}{\textbf{Injection}}&\multicolumn{2}{c}{\textbf{Shift}}\\
\cline{3-8}
Response & Transition & L &C & L& C & L & C\\
\hline
\multirow{2}{*}{Photoconductivity}
    & Delocalized & \cmark & \cmark$^*$ &\cmark$^*$ & \cmark&\cmark&\cmark$^*$ \\
   & Localized  &\xmark &\xmark &\xmark & \xmark&\cmark&\cmark$^*$\\
\hline\hline
\end{tabular}
\caption{Summary of the third order photoconductivity for localized and delocalized optical transitions. Here L/C denotes linearly/circularly polarized light and \cmark $\,$ denotes allowed and \xmark  $\,$ denotes forbidden. Injection and shift responses can be either longitudinal or Hall, whereas jerk responses are longitudinal. Entries marked with an asterisk require broken $T$ or $PT$ symmetry.}
\label{tab:light_polarization}
\end{table}

We now discuss the general properties of the photoconductivities. First, the jerk current formula is symmetric in $ab$ and therefore does not support Hall currents. By contrast, the injection and shift current generally allow for both longitudinal and Hall responses. In a $T$- or $PT$-symmetric system, the third-order injection current is anti-symmetric under $c\leftrightarrow d$, whereas the third-order shift current is symmetric under $c\leftrightarrow d$. Consequently, in $T$- or $PT$-symmetric systems the third-order injection current is induced only by circularly polarized light (CPL) and vanishes for linearly polarized light (LPL); similarly, the third-order shift current is induced only by LPL and vanishes for CPL in $T$- or $PT$-symmetric systems. 

These general considerations are qualitatively modified for localized optical transitions such as excitons. In particular, the third-order excitonic jerk and injection currents vanish identically, leading to striking consequences. For example, in $T$- or $PT$-symmetric systems, no excitonic current is generated under CPL, while under LPL the excitonic current is purely of shift-current origin. This contrasts sharply with the delocalized case, where the response is dominated by the injection current under CPL and by the jerk current (the longitudinal component) under LPL. These symmetry properties and the corresponding distinction between localized and delocalized optical transitions are summarized in Table.~\ref{tab:light_polarization}.

\textcolor{blue}{{\it Third Order Photoconductivity in MoS$_2$.}}
Finally, we illustrate the contrasting behavior between localized and delocalized optical transitions by computing the third order photoconductivity in MoS$_2$ focusing on the injection and shift current for excitonic and free delocalized optical transitions. We study a three-band tight-binding model for monolayer MoS$_2$~\cite{liu2013three} with screened Coulomb potential~\cite{keldysh2024coulomb,cudazzo2011} and tunable strain~\cite{rostami2015theory} along the armchair direction. 

\begin{figure}
    \centering
\includegraphics[width=1\linewidth]{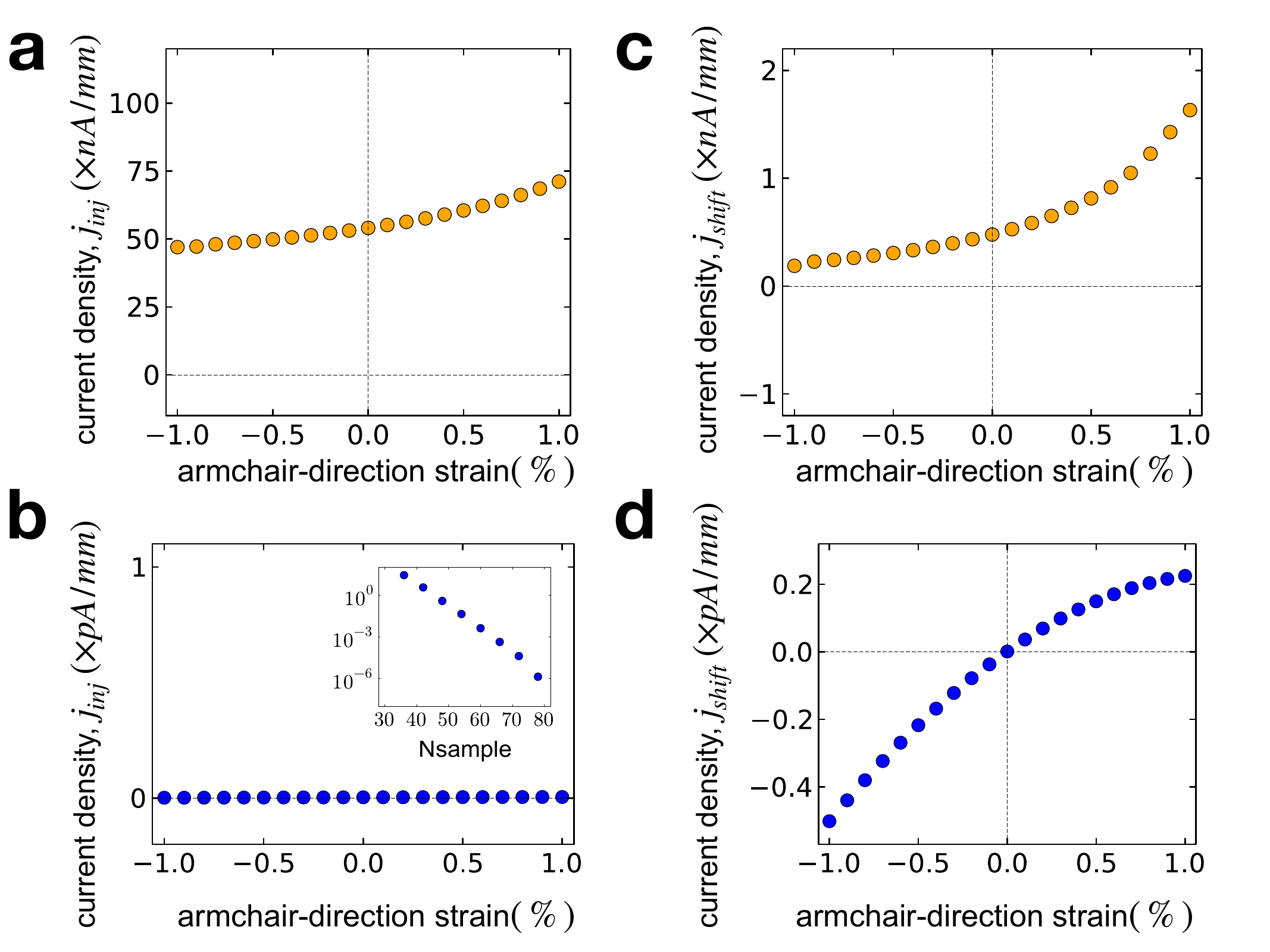}
    \caption{\textcolor{blue}{\it Third-order injection and shift photoconductive currents in monolayer MoS$_2$.} 
    {\bf a.} Third-order CPL injection Hall photoconductive current for delocalized free electronic optical transitions. The response remains finite over the strain range and measures $\sigma_{\text{inj}}^{[x,y][x,y]}$, obtained by antisymmetrizing over the first two and last two indices. {\bf b.} Excitonic CPL third-order injection Hall photoconductive current with photon energy corresponding to the lowest exciton energy. The response is numerically zero on a $60\times 60$ lattice. Inset: the finite size scaling of excitonic third order injection current at a fixed strain $0.5\%$, showing exponential decay. {\bf c.} Third-order longitudinal shift current for delocalized free electronic optical transitions under LPL (both along the armchair direction); this response was computed following Ref.~\cite{fregoso2019bulk, fregoso2020erratum}, see detailed discussion in {\bf SI}. This measures $\sigma^{yyyy}_{\rm shift}$. The response remains finite but is about two orders of magnitude smaller than the non-interacting injection current shown in panel (a). {\bf d.} Excitonic third-order longitudinal shift current under LPL at a photon energy $10$ meV above the lowest exciton energy to ensure a non-negligible contribution from $\partial_{\omega}\delta(\omega-\omega_{n0})$ term; here the response was computed using Eq.~(\ref{eq:shift_exciton_formula}). The response is finite for nonzero strain and vanishes at zero strain due to $C_{3z}$ symmetry. Since the excitonic jerk and injection currents vanish, this shift current is the leading third-order excitonic photoconductive current. See \textbf{SI} for details of the numerical simulation. In panel {\bf a,c} photon energy of $\hbar\omega=2.6$ eV was used above the optical gap; injection responses were computed using Eq.~(\ref{eq:injectionmain}). 
    }
    \label{fig:fig2}
\end{figure}

We first examine the injection current contribution to the third order photoconductivity. For the third order injection current (photoconductivity), we focus on the photovoltaic Hall effect under CPL illumination (notice that LPL contribution vanishes for third order injection currents in $T$-symmetric materials, see Table.~\ref{tab:light_polarization}). For delocalized free electrons, we find large third-order injection currents (photoconductivity) at photon energy $\hbar\omega=2.6$ eV see Fig.~\ref{fig:fig2}(a); this high photon energy is above the optical gap in the particle-hole continuum of MoS$_2$. The third-order injection currents in this case can be continuously tuned by varying strain as seen in Fig.~\ref{fig:fig2}(a). 

In stark contrast, we find the excitonic optical transition's third order injection current {\it vanishes}, see Fig.~\ref{fig:fig2}(b). This agrees with our analysis above and arises from the flat and uniform quantum geometry of resonant excitonic transitions as described in Eq.~\ref{eq:K_and_Omega} and Table.~\ref{tab:resonant_qg}. To further substantiate this result, we performed a finite-size scaling analysis of excitonic third order injection Hall current at a fixed strain $0.5\%$ (Fig.~\ref{fig:fig2}(b) inset), showing an exponential decay with system size. This provides compelling numerical evidence that the excitonic third order injection current vanishes in the thermodynamic limit. Notice that since injection currents are the only third order circular photoconductivity responses (see Table.~\ref{tab:light_polarization}), we find that such excitonic optical transitions have zero third-order circular photoconductivity. 

We now turn to numerical simulations of the shift current. Here we focus on the longitudinal current under linearly polarized light (both along the armchair direction); notice that CPL contribution vanishes for third order shift currents in $T$-symmetric materials, see Table.~\ref{tab:light_polarization}. We find the third-order shift currents for non-interacting electrons at $\hbar\omega=2.6$ eV is sizeable and exhibits a strain tunable magnitude. In contrast to the excitonic injection response, the excitonic shift current, evaluated using Eq.~\eqref{eq:shift_exciton_formula}, remains finite once strain is applied (Fig.~\ref{fig:fig2}(d)). It vanishes at zero strain due to the $C_{3z}$ rotational symmetry; a detailed discussion of this property is presented in the \textbf{SI}. Although the excitonic shift current is possibly several orders of magnitude smaller than its non-interacting counterpart, it remains the leading third-order excitonic photocurrent since both excitonic jerk and injection contributions vanish.

We have shown that interactions dramatically reshape the resonant nonlinear optical response of bound particle-hole excitations by flattening their quantum geometry: Hermitian curvature $K$ vanishes; metric $Q$ and connection $C$ are uniform. This is in sharp contrast to delocalized optical transitions that exhibit a non-vanishing and highly structured Hermitian curvature $K$ and non-uniform metric $Q$ and connection $C$. The interaction-induced flat quantum geometry we find in turn sharply constrain the nonlinear response of localized optical transitions (such as excitons): zeroing its third order circular photoconductivity in non-magnetic materials; this also locks their Hall response to that of the ground state. 

This interaction-induced flattening is particularly striking since it demonstrates interaction-driven departures from a simple non-interacting phenomenology for nonlinear optical response. While we have focused on the third order excitonic photoconductivity and its relation to the interacting quantum geometry of localized optical transitions, our many-body length gauge formulation is general. We anticipate that a range of other interaction-driven unconventional responses may similarly arise either from exotic ground states \cite{cai2023signatures,park2023observation,dong2024anomalous} or interacting many-particle optical excitations \cite{macdonald1988t, bonsall1977some,song2016chiral,kumar2016chiral,kaneko2021bulk,arora2022quantum}. In this regard, the many-body length gauge formulation we use provides a systematic methodology capable of incorporating a range of advanced many-body numerical methods for estimating interacting states \cite{sandvik2010computational} and relating these to a growing rich landscape of many-body quantum geometry.

\textcolor{blue}{\it Acknowledgments.} This research is supported by the Ministry of Education, Singapore, under its Academic Research Fund Tier 2 grant MOE-T2EP50225-0008 and Academic Research Fund Tier 3 grant MOE-MOET32023-0003 “Quantum Geometric Advantage”.

\bibliographystyle{apsrev4-2}
\bibliography{exciton}

\clearpage

\setcounter{equation}{0}             
\renewcommand{\theequation}{S\arabic{equation}}

\begin{widetext}

\section*{Supplementary Information}

\section{Evaluating the Hermitian connection for a Slater determinant ground state}

In this section we evaluate the Berry connection part of the Hermitian connection in the main text for a Slater determinant ground state. Below, all states and wave-functions depend on $\bm{\kappa}$, which we suppress for notational simplicity.

The Slater determinant ground state is $\ket{\Phi_0}=\prod_{a\in\rm{occ.}}c_a^{\dagger}\ket{0}$, where $c_a^{\dagger}=\int d\bm{x}\, \phi_a(\bm{x})a_{\bm{x}}^{\dagger}$ creates an electron with wave-function $\phi_a$. Conversely we have $a^{\dagger}_{\bm{x}}=\sum\limits_{a}\phi_a^*(\bm{x})c^{\dagger}_a$. Wick contraction yields: $\bra{\Phi_0}c^{\dagger}_ac_b\ket{\Phi_0}=\delta_{a,b}$ for occupied orbitals, and $\bra{\Phi_0}c_ic^{\dagger}_{j}\ket{\Phi_0}=\delta_{i,j}$ for unoccupied orbitals.

A particle-hole excitation removes an electron from an occupied orbital and put it in an empty orbital. Therefore $\ket{\Phi_n}$ can be expanded as:
\begin{align}
\ket{\Phi_n}=\sum\limits_{j\in\rm{unocc.},a\in\rm{occ.}}Z_{ja}c^{\dagger}_jc_a\ket{\Phi_0}.
\end{align}

Wick theorem then gives $\Psi(\bm{x}_e,\bm{x}_h)\equiv \bra{\Phi_0}a_{\bm{x}_h}^{\dagger}a_{\bm{x}_e}\ket{\Phi_n}=\sum\limits_{j\in \rm{unocc.},a\in\rm{occ.}}Z_{ja}\phi_j(\bm{x}_e)\phi_a^*(\bm{x}_h)$. Using the relation between $c$ and $a$, $\ket{\Phi_n}$ can equivalently be written as $\ket{\Phi_n}=\int d\bm{x}_e\,d\bm{x}_h\Psi(\bm{x}_e,\bm{x}_h)a^{\dagger}_{\bm{x}_e}a_{\bm{x}_h}\ket{\Phi_0}.$

We can now use this representation to evaluate the Berry connection under flux threading. Differentiating $\ket{\Phi_n}$ with respect to $\bm{\kappa}$, we get:
\begin{align}\label{eq:phi_n_derivative}
i\partial_{\bm{\kappa}}\ket{\Phi_n}=\int d\bm{x}_e\,d\bm{x}_h\,\big(i\partial_{\bm{\kappa}}\Psi(\bm{x}_e,\bm{x}_h)\big)a^{\dagger}_{\bm{x}_e}a_{\bm{x}_h}\ket{\Phi_0}+i\int d\bm{x}_e\,d\bm{x}_h\,\Psi(\bm{x}_e,\bm{x}_h)a^{\dagger}_{\bm{x}_e}a_{\bm{x}_h}\ket{\partial_{\bm{\kappa}}\Phi_0}.
\end{align}

For the first term on the RHS of Eq.~\ref{eq:phi_n_derivative} after taking the inner product with $\bra{\Phi_n}$, we use the definition of $\Psi$, $\bra{\Phi_n}a^{\dagger}_{\bm{x}_e}a_{\bm{x}_h}\ket{\Phi_0}=\Psi^*(\bm{x}_e,\bm{x}_h)$ to obtain $i\int d\bm{x}_e d\bm{x}_h \Psi^*(\bm{x}_e,\bm{x}_h)  \partial_{\bm{\kappa}}\Psi(\bm{x}_e,\bm{x}_h)$.

For the second term on the RHS of Eq.~\ref{eq:phi_n_derivative}, we use the fact that $\partial_{\bm{\kappa}}\ket{\Phi_0}$ acts on one orbital at a time, and contains at most a single particle-hole excitation. Hence $i\partial_{\bm{\kappa}}\ket{\Phi_0}=\mathcal{\bm{A}}_0\ket{\Phi_0}+\sum\limits_{j\in\rm{unocc.},a\in\rm{occ.}}Y_{ja}c^{\dagger}_jc_a\ket{\Phi_0}$. Defining $\hat{X}=\int d\bm{x}_e\,d\bm{x}_h\,\Psi(\bm{x}_e,\bm{x}_h)a^{\dagger}_{\bm{x}_e}a_{\bm{x}_h}=\sum\limits_{j\in\rm{unocc.},a\in\rm{occ.}}Z_{ja}c^{\dagger}_jc_a$. We note that $\hat{X}c^{\dagger}_jc_a\ket{\Phi_{0}}$ is either zero or contains two particle-hole pairs relative to $\ket{\Phi_0}$ and is therefore orthogonal to $\ket{\Phi_n}$ containing only single particle-hole excitation. The only surviving contribution is therefore $\bra{\Phi_0}i\partial_{\bm{\kappa}}\ket{\Phi_0}$, which cancels $\mathcal{\bm{A}}_0$. In summary, after restoring $\bm{\kappa}$ dependence, we have:
\begin{align}
\mathcal{\bm{A}}_n-\mathcal{\bm{A}}_0=i\int d\bm{x}_e d\bm{x}_h \Psi^*_{\bm{\kappa}}(\bm{x}_e,\bm{x}_h)  \partial_{\bm{\kappa}}\Psi_{\bm{\kappa}}(\bm{x}_e,\bm{x}_h).
\end{align}

\section{Connecting the finite-flux response to zero-flux response in the many-body length gauge}
We now justify replacing the $\bm{A}$-dependent quantities in the length-gauge equation Eq.~\ref{eq:rho^E_recursive} by their zero-flux counterparts for an insulator. Without loss of generality, we impose periodic boundary condition along the $x$ direction with linear size $L$. We first show that the physical response is invariant under insertion of an integer number of flux quanta. To this end, we take $A=2\pi q/L$, with $q$ an integer. Gauge invariance\cite{kohn1964theory,resta1998quantum} ensures:
$UH(2\pi q/L)U^{-1}=H(0)$, with $U=\exp\left(i\frac{2\pi q}{L}\sum_i x_i\right)$ being the large gauge transformation. Adiabatically following an eigenstate $\ket{\Phi_n(A)}$ from $A=0$ to $2\pi q/L$ and applying $U$ therefore maps it to an eigenstate of the original Hamiltonian. In general, this state need not be the initial eigenstate, giving rise to spectral flow:
\begin{equation}\label{eq:stitch}
U\ket{\Phi_n(2\pi q/L)}=\ket{\Phi_{\pi(n)}(0)}.
\end{equation}
Here $\pi$ is a permutation of the many-body eigenstates. It immediately follows that the eigen-energy satisfies $E_n(2\pi q/L)=E_{\pi(n)}(0).$ For an insulator with a unique ground state, $\pi(0)=0$. Introducing the stitch matrix
$\Pi_{nm}=\delta_{\pi(n),m},$
the zeroth-order density matrix therefore satisfies $
\rho^{E,(0)}(2\pi q/L)=\Pi\rho^{E,(0)}(0)\Pi^{-1}.$

Using Eq.~\ref{eq:stitch}, we find that quantities entering the recursive length-gauge equation Eq.~\ref{eq:rho^E_recursive} satisfies:
\begin{equation}
\omega_{nm}(2\pi q/L)=\omega_{\pi(n)\pi(m)}(0), \qquad \bm{\xi}_{nm}(2\pi q/L)=\bm{\xi}_{\pi(n)\pi(m)}(0).
\end{equation}
The recursion relation is thus covariant under $\Pi$, and to arbitrary order in $\bm{E}$ we have: $\rho^E(2\pi q/L)=\Pi\rho^E(0)\Pi^{-1}.$ Similarly, $O^E(2\pi q/L)=\Pi O^E(0)\Pi^{-1},$ so that (using the shorthand $\braket{O}_A=\text{Tr}\big[O^E(A)\rho^E(A)\big]$):
\begin{equation}
\braket{O}_{2\pi q/L}=\braket{O}_0.
\end{equation}

For a general $A$, we can write $A=\frac{2\pi q}{L}+\alpha_L$ with $|\alpha_L|<\frac{2\pi}{L}$, so that $\alpha_L\rightarrow 0$ when $L\rightarrow\infty$. Using gauge invariance to remove the integer-flux part in $A$, and noting that physical response has a well-defined thermodynamic limit, we have $\text{lim}_{L\rightarrow \infty}\braket{O}_A=\text{lim}_{L\rightarrow\infty}\braket{O}_{\alpha_L}=\text{lim}_{L\rightarrow\infty}\braket{O}_0$. Thus the finite-A and zero-flux responses are identical in the thermodynamic limit, justifying the use of many-body length-gauge equation evaluated at zero flux.

\section{Simplification of the general expression}
The many-body length-gauge Liouville equation is:
\begin{align}
   [i\partial_t-\omega_{nm}]\rho^E_{nm}=i\frac{e}{\hbar}\bm{E}\cdot[D_{\bm{\kappa}}\rho^E]_{nm},
\end{align}
where $[D_a\rho]_{nm}\equiv \rho_{nm;a}-i[r^a,\rho]_{nm}$ with $\bm{r}_{nm}=(1-\delta_{nm})\bm{\xi}_{nm}$, and $O_{nm;a}\equiv\partial_aO_{nm}-i(\mathcal{A}_n^a-\mathcal{A}^a_m)O_{nm}$ denoting the generalized derivative. The general photocurrent expression is then: $j^a=-e\text{Tr}(v^a\rho^{(l)}),$ where $\rho^{(l)}$ is solved from the Liouville equation in the length gauge. For simplicity, we set $e=\hbar=1$ throughout the calculation and restore the overall factor $\frac{e^4}{\hbar^3V}$ at the end, appropriate for the third-order current density.

Let's simplify the photocurrent expression to classify the full response into terms with different divergence properties.

First, notice that the full current has time dependence $e^{-i\omega_{\Sigma} t}$. Therefore $\rho^{(l)}_{nm}=\frac{iE^b[D_{b}{\rho^{(l-1)}}]_{nm}}{\omega_{\Sigma}-\omega_{nm}}$. Next, we decompose the current into contributions arising from the diagonal and off-diagonal elements of $\rho^{(l)}$, i.e., $\bm{j}=\bm{j}_{\rm diag}+\bm{j}_{\rm off-diag}$, with:
\begin{align}
& j^a_{\rm diag}=\frac{-i}{\omega_{\Sigma}}\sum\limits_{n}v^a_{nn}E^b[D_b\rho^{(l-1)}]_{nn},\\
& j^a_{\rm off-diag}=\sum\limits_{n\neq m}\frac{-i}{(\omega_{\Sigma}-\omega_{nm})}v^a_{mn}E^b[D_b\rho^{(l-1)}]_{nm}=\sum\limits_{n, m}\frac{\omega_{mn}}{(\omega_{\Sigma}-\omega_{nm})}r^a_{mn}E^b[D_b\rho^{(l-1)}]_{nm},
\end{align}
where we have used $v^a_{mn}=i\omega_{mn}r^a_{mn}$ and have lifted the constraint $n\neq m$. We can further massage $\bm{j}_{\rm off-diag}$ by noticing that $\frac{\omega_{mn}}{\omega_{\Sigma}-\omega_{nm}}=1-\frac{\omega_{\Sigma}}{\omega_{\Sigma}-\omega_{nm}}$. Therefore $\bm{j}_{\rm off-diag}$ is composed of two parts $\bm{j}_{\rm off-diag}=\bm{j}^{(1)}_{\rm off-diag}+\bm{j}^{(2)}_{\rm off-diag}$, with:
\begin{align}
&j^{a,(1)}_{\rm off-diag}=\sum\limits_{n,m}\frac{-\omega_{\Sigma}}{\omega_{\Sigma}-\omega_{nm}}r^a_{mn}E^b[D_b\rho^{(l-1)}]_{nm}=i\omega_{\Sigma}\text{Tr}(r^a\rho^{(l)}),\\
&j^{a,(2)}_{\rm off-diag}=\sum\limits_{n,m}r^a_{mn}E^b[D_b\rho^{(l-1)}]_{nm}=-\sum\limits_{n, m}E^b[D_br^a]_{mn}\rho^{(l-1)}_{nm},
\end{align}
where, in the second line, we have used the cyclic property of the trace and the fact that the full expression $\mathrm{Tr}(r\rho^{(l-1)})$ is independent of $\kappa$. The current $\bm{j}^{(1)}_{\rm off-diag}$ vanishes in the DC limit $\omega_{\Sigma}\rightarrow 0$ and will be omitted below. The current $\bm{j}^{(2)}_{\rm off-diag}$ can be further simplified by explicitly evaluating $[D_br^a]_{mn}$. The following identity will be useful: $\partial_b(r^a_{mn})-\partial_a(r^b_{mn})=i\braket{\partial_bm|\partial_an}-i\braket{\partial_am|\partial_bn}=i[\xi^b,\xi^a]_{mn}$. Then, for $m\neq n$,
\begin{align}
& [D_br^a]_{mn}=\partial_b(r^a_{mn})-i[\xi^b,r^a]_{mn}=\partial_a(r^b_{mn})+i[\xi^b,\xi^a]_{mn}-i[\xi^b,r^a]_{mn}=\partial_a(r^b_{mn})+ir^b_{mn}(\xi^a_{nn}-\xi^a_{mm})=r^b_{mn;a}.
\end{align}

And for $m=n$, we have $[D_br^a]_{nn}=-i[\xi^b,r^a]_{nn}=\partial_a\xi^b_{nn}-\partial_b\xi^a_{nn},$ which corresponds to the anomalous velocity.

Finally, collecting $\bm{j}_{\rm diag}$ and $\bm{j}^{(2)}_{\rm off-diag}$, we arrive at:
\begin{align}\label{eq:general_current_expression}
j^a=\frac{1}{i\omega_{\Sigma}}\sum\limits_{n}v^a_{nn}E^b[D_b\rho^{(l-1)}]_{nn}-\sum\limits_{n\neq m}E^br^b_{mn;a}\rho^{(l-1)}_{nm}-\sum\limits_nE^b\Omega^{ab}_n\rho^{(l-1)}_{nn}
\end{align}

As suggested from their $\omega_{\Sigma}$ dependence, these terms contribute to different parts of responses. Let's consider $DC$ current with $\omega_{\Sigma}\rightarrow 0$. As a simple check, we look at the case of the second order current with $l=2$, the first term gives rise to injection current, the second term gives rise to the shift current and the anomalous velocity term vanishes since $\rho^{(1)}$ is not diagonal. With the general expression we turn to the more complicated case of third-order response in the next section.

\section{Derivation of the third-order response: preliminary}

To obtain the third-order response, we insert $\rho^{(2)}$ into Eq.~\eqref{eq:general_current_expression}. Below we work out the density matrix up to second order, with subscripts indicating spatial components of light-matter interactions. The first order $\rho^{(1)}_{nm}$ is obtained by $\rho^{(1)}_{d,nm}=\frac{iE^d[D_d\rho^{(0)}]_{nm}}{\omega_d-\omega_{nm}}=\frac{E^d[r^d,\rho^{(0)}]_{nm}}{\omega_d-\omega_{nm}}$, which is non-zero only when $n$ or $m$ is equal to 0:
\begin{align}
&\rho^{(1)}_{d,n0}=\frac{E^dr^d_{n0}}{\omega_d-\omega_{n0}},\qquad \rho^{(1)}_{d,0n}=\frac{-E^dr^d_{0n}}{\omega_d-\omega_{0n}}.
\end{align}

The second order $\rho^{(2)}$ is obtained by: $\rho^{(2)}_{cd,nm}=\frac{iE^c[D_c\rho^{(1)}_d]_{nm}}{\omega_c+\omega_d-\omega_{nm}}$, which has a commutator piece $[r^c,\rho^{(1)}_d]_{nm}$, and a derivative piece $(\rho^{(1)}_{d,nm})_{;c}$ only when $n$ or $m$ is equal to 0. Below we list terms that are used in calculation with $n,m\neq 0$:
\begin{align}
&\rho^{(2)}_{cd,n0}=\frac{iE^c(\rho^{(1)}_{d,n0})_{;c}}{\omega_c+\omega_d-\omega_{n0}}+\frac{E^c[r^c,\rho^{(1)}_d]_{n0}}{\omega_c+\omega_d-\omega_{n0}}, \qquad \rho^{(2)}_{cd,0n}=\frac{iE^c(\rho^{(1)}_{d,0n})_{;c}}{\omega_c+\omega_d-\omega_{0n}}+\frac{E^c[r^c,\rho^{(1)}_d]_{0n}}{\omega_c+\omega_d-\omega_{0n}}\\
&\rho^{(2)}_{cd,nn}=\frac{E^c(r^c_{n0}\rho^{(1)}_{d,0n}-\rho^{(1)}_{d,n0}r^c_{0n})}{\omega_c+\omega_d},\qquad \rho^{(2)}_{cd,00}=\sum\limits_{n\neq 0}\frac{E^c(r^c_{0n}\rho^{(1)}_{d,n0}-\rho^{(1)}_{d,0n}r^c_{n0})}{\omega_c+\omega_d}=-\sum\limits_n\rho^{(2)}_{cd,nn},\\
&\rho^{(2)}_{cd,nm}=\frac{E^c(r^c_{n0}\rho^{(1)}_{d,0m}-\rho^{(1)}_{d,n0}r^c_{0m})}{\omega_c+\omega_d-\omega_{nm}}.
\end{align}

Now let's further massage Eq.~\eqref{eq:general_current_expression} at third order. Notice that $\rho^{(2)}_{00}=-\sum\limits_n\rho^{(2)}_{nn}$, and expand $[D\rho^{(2)}]$ into generalized derivative and commutator term, we can further simplify it to be (note $v^a_{nn}-v^a_{00}=\partial_a(\omega_{n0})$):
\begin{align}
j^a=\sum\limits_{\text{perm of }bcd}    \Bigg(
&\underbrace{\frac{1}{i\omega_{\Sigma}}\sum_{n} \partial_a(\omega_{n0}) E^b (\rho^{(2)}_{cd,nn})_{;b}}_{\text{diagonal-derivative term}} \label{eq:diag_derivative}\\
+&\underbrace{\frac{(-i)}{i\omega_{\Sigma}}\sum\limits_{n m}E^b\partial_a(\omega_{nm})r^b_{nm}\rho^{(2)}_{cd,mn}}_{\text{diagonal-commutator term}} \label{eq:diag_comm}\\
&- \underbrace{\sum_{n\neq m} E^b r^b_{mn;a} \rho^{(2)}_{cd,nm}}_{\text{interband term}}\label{eq:third_order_current_inter} \\
&- \underbrace{\sum_{n} E^b \Omega^{ab}_{n0} \rho^{(2)}_{cd,nn}}_{\text{anomalous vel. term}}
\Bigg).\label{eq:third_order_current_anom}
\end{align}

The third-order response is classified into three different terms, first is the jerk current that is proportional to $\frac{1}{\omega_{\Sigma}^2}$, second is the injection current that is proportional to $\frac{1}{\omega_{\Sigma}}$, and third is the shift current of order $(\omega_{\Sigma})^0$. In practice, these $1/ \omega_{\Sigma}$ divergences are regularized by relaxation time with $-i\omega_\Sigma \to 1/\tau$. In the main text, we therefore group the jerk, injection, and shift terms according to their $\tau$ dependencies. Throughout the appendices, however, we keep the $\omega_\Sigma$ explicitly to make contact with Ref.~\cite{fregoso2019bulk}, .

\section{Jerk current response}
The jerk current response originates from the diagonal-derivative term Eq.~\eqref{eq:diag_derivative} via $\rho^{(2)}_{cd,nn}$ and $\rho^{(2)}_{dc,nn}$, which has a potential $\frac{1}{\omega_{\Sigma}}$ divergence to account for the full $(\omega_{\Sigma})^{-2}$ divergence. Therefore we have:
\begin{align}
&j^a_{\text{jerk}}=\sum\limits_n\frac{1}{i\omega_{\Sigma}}\partial_a(\omega_{n0})E^b\partial_b\big(\rho^{(2)}_{cd,nn}+\rho^{(2)}_{dc,nn}\big)=\sum\limits_n\frac{E^bE^cE^d}{i\omega_{\Sigma}^2}\partial_a\partial_b(\omega_{n0})\big[(\frac{1}{\omega_d-\omega_{n0}}+\frac{1}{\omega_c-\omega_{0n}})r^d_{n0}r^c_{0n}+(c\leftrightarrow d)\big]\nonumber\\
&=\frac{2\pi e^4}{\hbar^3V}\frac{E^bE^cE^d}{(-i\omega_{\Sigma})^2}\sum\limits_{n} \partial_a\partial_b(\omega_{n0})r^d_{n0}r^c_{0n}\delta(\omega_{n0}-\omega),
\end{align}
where we have shifted $\partial_b$ around in the first line, and have restored the physical constants in the end.
\section{Injection current response}
The injection current response is of order $\frac{1}{\omega_{\Sigma}}$. It can be arranged into the following terms:
\begin{enumerate}
    \item In the diagonal-commutator contribution $[r^b,\rho^{(2)}_{cd}]_{nn}$ Eq.~\eqref{eq:diag_comm}, $\rho^{(2)}_{cd}$ contains a commutator of the form $(r^c_{n0}\rho^{(1)}_{d,0m}-\rho^{(1)}_{d,n0}r^c_{0m})/(\omega_c+\omega_d-\omega_{nm})$. Substituting this back and summing over permutations generates a contribution of the form $\partial_a(\omega_{n0})r^c_{0n}\frac{r^b_{nm}}{\omega_{nm}}r^d_{m0}$ involving virtual transitions to off-resonant bands.
    \item In the diagonal-commutator term $[r^b,\rho^{(2)}_{cd}]_{nn}$ Eq.~\eqref{eq:diag_comm}, $\rho^{(2)}_{cd}$ contains a derivative piece $\nabla_c(\rho^{(1)}_{d,n0})/(\omega_c+\omega_d-\omega_{n0})$. This gives rise to two geometric terms, one is of the form $(r^c_{0n;b}r^d_{n0})_{;a}$, and one is of the form $\partial_a(\omega_{n0})(\frac{r^c_{0n}}{\omega_{0n}})_{;b}r^d_{n0}$.
    \item The anomalous velocity term Eq.~\eqref{eq:third_order_current_anom} gives rise to a term with $\frac{1}{\omega_{\Sigma}}$ divergence, which originates from $\rho^{(2)}_{cd,nn}$.
\end{enumerate}

\subsection{Pure commutators in $\rho^{(2)}_{nm}$ from the diagonal-commutator part Eq.~\eqref{eq:diag_comm}}
In this section we will focus on the pure commutator contribution to $\rho^{(2)}$ in Eq.~\eqref{eq:diag_comm}. Below we will use a shorthand $\mathcal{C}_1=\frac{1}{-i\omega_{\Sigma}}E^bE^cE^d$. 

For $n,m\neq 0$, we can express $\rho^{(2)}_{nm}$ in terms of $-i[r,\rho^{(1)}]_{nm}$ and obtain:
\begin{align}
&-i\mathcal{C}_1\sum\limits_{\{ bcd\},nm}\partial_a(\omega_{nm})\frac{r^b_{nm}}{(\omega_c+\omega_d-\omega_{mn})}\big[\frac{r^c_{m0}r^d_{0n}}{\omega_d-\omega_{0n}}+\frac{r^d_{m0}r^c_{0n}}{\omega_d-\omega_{m0}}\big]\nonumber\\
& =-i\mathcal{C}_1 \sum\limits_{\{ bcd\},nm}\partial_a(\omega_{nm})\frac{r^b_{nm}r^c_{m0}r^d_{0n}}{(\omega_c+\omega_d-\omega_{mn})} \big(\frac{1}{\omega_d-\omega_{0n}}+\frac{1}{\omega_c-\omega_{m0}}\big)=(-i)\mathcal{C}_1 \sum\limits_{\{ bcd\},nm}\partial_a(\omega_{nm})\frac{r^b_{nm}r^c_{m0}r^d_{0n}}{(\omega_d-\omega_{0n})(\omega_c-\omega_{m0})}\nonumber\\
&= -i\mathcal{C}_1\sum\limits_{\{bcd\},nm}
\partial_a(\omega_{n0})
\frac{r^b_{nm}r^c_{m0}r^d_{0n}}{(\omega_d-\omega_{0n})(\omega_c-\omega_{m0})}\qquad  \text{Term (I)}\label{eq:term_1}\\
&+(-i)\mathcal{C}_1\sum\limits_{\{bcd\},nm}
\partial_a(\omega_{0m})
\frac{r^b_{nm}r^c_{m0}r^d_{0n}}{(\omega_d-\omega_{0n})(\omega_c-\omega_{m0})} \qquad\text{Term (II)}\label{eq:term_2}
\end{align}

For $m=0$ (and renaming $n$ with $m$, and exchanging $b\leftrightarrow c$), we get Term (III):
\begin{align}\label{eq:term_3}
i\mathcal{C}_1\sum\limits_{\{bcd\},nm}\partial_a(\omega_{m0})\frac{r^b_{nm}r^c_{m0}r^d_{0n}}{(\omega_b+\omega_d-\omega_{0m})(\omega_d-\omega_{0n})} \qquad \text{Term (III)}.
\end{align}

Adding Term (II) and Term (III), we obtain the resonant contribution:
\begin{align}
    \text{(II)+(III)}=2\pi\mathcal{C}_1\sum\limits_{\{bcd\},nm}\partial_a(\omega_{m0})r^b_{nm}r^c_{m0}r^d_{0n}\frac{1}{\omega_d-\omega_{0n}}\delta(\omega_c-\omega_{m0}).
\end{align}

After summing over all permutations, it gives two terms:
\begin{align}
2\pi\mathcal{C}_1\sum\limits_{nm}\delta(\omega-\omega_{m0})\partial_a(\omega_{m0}) \big(\frac{r^b_{nm}r^d_{m0}r^c_{0n}}{\omega_{nm}}-\frac{r^c_{nm}r^d_{m0}r^b_{0n}}{\omega_{0n}}\big)
\end{align}

And we also have a Term (IV) by setting $n$ to $0$,  renaming $m$ by $n$ and permuting $(cdb)\rightarrow (bcd)$:
\begin{align}\label{eq:term_4}
i\mathcal{C}_1\sum\limits_{\{bcd\},nm}\partial_a(\omega_{0n})\frac{r^b_{nm}r^c_{m0}r^d_{0n}}{(\omega_b+\omega_c-\omega_{n0})(\omega_c-\omega_{m0})} \qquad \text{Term (IV)}.
\end{align}

Adding Term (I) and Term (IV), we have the resonant part from $\frac{1}{\omega_d-\omega_{0n}}$:
\begin{align}
    \text{(I)+(IV)}=-2\pi\mathcal{C}_1\sum\limits_{\{bcd\},nm}\partial_a(\omega_{n0})r^b_{nm}r^c_{m0}r^d_{0n}\frac{1}{\omega_c-\omega_{m0}}\delta(\omega_d-\omega_{0n}).
\end{align}

This yields another two terms after summing over permutations:
\begin{align}
2\pi\mathcal{C}_1\sum\limits_{nm}\delta(\omega-\omega_{n0})\partial_a(\omega_{n0}) \big(\frac{r^d_{nm}r^b_{m0}r^c_{0n}}{\omega_{m0}}-\frac{r^b_{nm}r^d_{m0}r^c_{0n}}{\omega_{nm}}\big)
\end{align}

In summary we have the following contribution to the injection response tensor:
\begin{align}
&j_{\text{inj-1}}^a=2\pi\mathcal{C}_1\sum_{n} \delta(\omega-\omega_{n0})\partial_a(\omega_{n0})\sum\limits_m\Big[\big(r^d_{nm}\frac{r^b_{m0}}{\omega_{m0}}-\frac{r^b_{nm}}{\omega_{nm}}r^d_{m0}\big)r^c_{0n}+r^d_{n0}\big(r^c_{0m}\frac{r^b_{mn}}{\omega_{mn}}-\frac{r^b_{0m}}{\omega_{0m}}r^c_{mn}\big)\big]
\end{align}

\subsection{$\nabla\rho^{(1)}_{n0}$ in $\rho^{(2)}_{n0}$ from the diagonal-commutator part Eq.~\eqref{eq:diag_comm} and remaining terms from the diagonal-derivative part Eq.~\eqref{eq:diag_derivative}}
The second part of injection current comes from derivative $\nabla \rho^{(1)}_{n0}$ to the diagonal-commutator Eq.~\eqref{eq:diag_comm}, and the remaining terms in the diagonal-derivative term Eq.~\eqref{eq:diag_derivative}:
\begin{align}
&j^a_{\text{inj}-2}=\frac{1}{i\omega_{\Sigma}}\sum\limits_{\{bcd\}}\sum\limits_{n}E^b\partial_a(\omega_{n0})\big[r^b_{n0}\frac{E^c}{\omega_c+\omega_d-\omega_{0n}}(\rho^{(1)}_{d,0n})_{;c}-r^b_{0n}\frac{E^c}{\omega_c+\omega_d-\omega_{n0}}(\rho^{(1)}_{d,n0})_{;c}\big]\nonumber\\
&+\frac{1}{i\omega_{\Sigma}}\big[\sum\limits_{n}\partial_a(\omega_{n0})E^c\nabla_c(\rho^{(2)}_{bd,nn}+\rho^{(2)}_{db,nn})+\sum\limits_{n}\partial_a(\omega_{n0})E^d\nabla_d(\rho^{(2)}_{bc,nn}+\rho^{(2)}_{cb,nn})\big].
\end{align}

Expanding them out, we have:
\begin{align}
&j^a_{\text{inj}-2}=\mathcal{C}_1\sum\limits_{\{bcd\}}\sum\limits_n\partial_a(\omega_{n0})\big[\frac{r^b_{n0}}{\omega_c+\omega_d-\omega_{0n}}(\frac{r^d_{0n}}{\omega_d-\omega_{0n}})_{;c}+\frac{r^d_{0n}}{\omega_c+\omega_b-\omega_{n0}}(\frac{r^b_{n0}}{\omega_b-\omega_{n0}})_{;c}\big]\nonumber\\
&+\mathcal{C}_1\sum\limits_n\partial_a(\omega_{n0})\Big\{\nabla_c\big[\frac{r^b_{n0}r^d_{0n}}{(\omega_d-\omega_{0n})(\omega_b-\omega_{n0})}+(b\leftrightarrow d)\big]+\nabla_d\big[\frac{r^b_{n0}r^c_{0n}}{(\omega_c-\omega_{0n})(\omega_b-\omega_{n0})}+(b\leftrightarrow c)\big]\Big\}
\end{align}

We expand the permutation and organize the terms into two separate groups. Terms in the first group include the $(dcb),(bcd),(cdb),(bdc)$ permutations from the diagonal commutative term and all the terms coming from the diagonal-derivative term, and are (we have simplified the expression using $\omega_b+\omega_c+\omega_d=0$):
\begin{align}
&\mathcal{C}_1\sum\limits_n\partial_a(\omega_{n0})\Big\{\Big[\big[\frac{r^d_{n0}}{-\omega_d-\omega_{0n}}(\frac{r^b_{0n}}{\omega_b-\omega_{0n}})_{;c}+\frac{r^b_{0n}}{-\omega_b-\omega_{n0}}(\frac{r^d_{n0}}{\omega_d-\omega_{n0}})_{;c}+\nabla_c\big(\frac{r^b_{0n}r^d_{n0}}{(\omega_b-\omega_{0n})(\omega_d-\omega_{n0})}\big)\big]+(b\leftrightarrow d)\Big]+ (c\leftrightarrow d)\Big\}.
\end{align}
It is clear from this expression that all the non-resonant contributions cancel with each other, therefore we are only left with resonant contributions, which are:
\begin{align}
&\mathcal{C}_1\sum\limits_n\partial_a(\omega_{n0})\big[r^d_{n0}(\frac{r^b_{0n}}{\omega_b-\omega_{0n}})_{;c}(\frac{1}{-\omega_d-\omega_{0n}}+\frac{1}{\omega_d-\omega_{n0}})+r^c_{0n}(\frac{r^b_{n0}}{\omega_b-\omega_{n0}})_{;d}(\frac{1}{-\omega_c-\omega_{n0}}+\frac{1}{\omega_c-\omega_{0n}})\big]\nonumber\\
&=2i\pi\mathcal{C}_1\sum\limits_n\partial_a(\omega_{n0})[r^d_{n0}(\frac{r^b_{0n}}{\omega_{0n}})_{;c}+r^c_{0n}(\frac{r^b_{n0}}{\omega_{n0}})_{;d}]\delta(\omega-\omega_{n0}).
\end{align}

Terms in the second group include the $(cbd),(dbc)$ permutations from the diagonal commutative term (i.e., covariant derivatives on $b$) requires extra care as multiple poles collide. Below we explicitly set $\omega_b=0$ below and only look at resonant contributions:
\begin{align}
&\mathcal{C}_1\sum\limits_n\partial_a(\omega_{n0})\big[\frac{r^d_{n0}}{\omega_c-\omega_{0n}}(\frac{r^c_{0n}}{\omega_c-\omega_{0n}})_{;b}+\frac{r^c_{0n}}{\omega_d-\omega_{n0}}(\frac{r^d_{n0}}{\omega_d-\omega_{n0}})_{;b}+(c\leftrightarrow d)\big]\nonumber\\
&=\mathcal{C}_1\sum\limits_n\partial_a(\omega_{n0})\big[\frac{r^d_{n0}(r^c_{0n})_{;b}}{(\omega_c-\omega_{0n})^2}+\frac{r^c_{0n}(r^d_{n0})_{;b}}{(\omega_d-\omega_{n0})^2}+(c\leftrightarrow d)\big],
\end{align}
where the triple poles $\propto \frac{\partial_b\omega_{0n}}{(\omega_c-\omega_{0n})^3}+\frac{\partial_b\omega_{n0}}{(\omega_d-\omega_{n0})^3}$ vanish because $(\frac{1}{\omega+i\eta})^3\propto \delta''(\omega)$, and $\delta''(x)=\delta''(-x)$. And the double pole can be massaged as follows. Notice that $\partial_a(\omega_{n0})\frac{1}{(\omega-\omega_{n0})^2}=\partial_a(\frac{1}{\omega-\omega_{n0}})$, and $\partial_a(\omega_{n0})\frac{1}{(-\omega-\omega_{0n})^2}=-\partial_a(\frac{1}{-\omega-\omega_{0n}})$, and shifting the $\partial_a$ around, we have (we have omitted the $c\leftrightarrow d$ term as it does not have resonant contributions):
\begin{align}
i\pi\mathcal{C}_1\sum\limits_n\delta(\omega-\omega_{n0})\partial_a[r^c_{0n}(r^d_{n0})_{;b}-r^d_{n0}(r^c_{0n})_{;b}].
\end{align}

In summary we have:
\begin{align}
j^a_{\text{inj}-2}=i\pi\mathcal{C}_1\sum\limits_n\delta(\omega-\omega_{n0})\Big[2\partial_a(\omega_{n0})\big[r^d_{n0}(\frac{r^b_{0n}}{\omega_{0n}})_{;c}+r^c_{0n}(\frac{r^b_{n0}}{\omega_{n0}})_{;d}\big]+\partial_a\big[r^c_{0n}(r^d_{n0})_{;b}-r^d_{n0}(r^c_{0n})_{;b}\big]\Big].
\end{align}

\subsection{Anomalous velocity term Eq.~\eqref{eq:third_order_current_anom}}
Finally we look at the anomalous velocity term:
\begin{align}
&j^{a}_{\text{inj-3}}=-E^b\sum\limits_n\Omega^{ab}_{n0}(\rho^{(2)}_{cd,nn}+\rho^{(2)}_{dc,nn})=\frac{E^bE^cE^d}{\omega_{\Sigma}}\sum\limits_n\Omega^{ab}_{n0}\big[(\frac{1}{\omega_d-\omega_{n0}}+\frac{1}{\omega_c-\omega_{0n}})r^d_{n0}r^c_{0n}+(c\leftrightarrow d)\big]\nonumber\\
&=-2\pi\mathcal{C}_1\sum\limits_n\delta(\omega_{n0}-\omega)\Omega^{ab}_{n0}r^c_{0n}r^d_{n0}.
\end{align}

\subsection{Summary of the third-order injection current}
In summary, the full expression for third-order many-body injection current is:
\begin{align}
&\sigma^{abcd}_{\rm inj}(\omega)=\sum_n\pi\delta(\omega_{n0}- \omega)\Bigg[2\partial_{a}(\omega_{n0})\sum\limits_{m}\big[\big(r^d_{nm}\frac{r^b_{m0}}{\omega_{m0}}-\frac{r^b_{nm}}{\omega_{nm}}r^d_{m0}\big)r^c_{0n}+r^d_{n0}\big(r^c_{0m}\frac{r^b_{mn}}{\omega_{mn}}-\frac{r^b_{0m}}{\omega_{0m}}r^c_{mn}\big)\big]\nonumber\\
&+2i\partial_a(\omega_{n0})\big[r^d_{n0}(\frac{r^b_{0n}}{\omega_{0n}})_{;c}+(\frac{r^b_{n0}}{\omega_{n0}})_{;d}r^c_{0n}\big]+i\partial_a\big[r^c_{0n}(r^d_{n0})_{;b}-r^d_{n0}(r^c_{0n})_{;b}\big]-2\Omega_{n0}^{ab}r^c_{0n}r^d_{n0}\Bigg].
\end{align} 

The first line involves virtual transition to off-resonant states. It can be identified with the field derivative of the Hermitian metric, $\partial_{E_b}Q^{n0}_{cd}$ under electric field $E^b$ if we denote: $\partial_{E_b}\ket{0}\equiv \frac{\bra{m}r^b\ket{0}}{\omega_{0m}}\ket{m}$ and $\partial_{E_b}\ket{n}\equiv \frac{\bra{m}r^b\ket{n}}{\omega_{nm}}\ket{m}$. Then we have $\partial_{E_b}Q^{n0}_{cd}=r^c_{0n}(-r^d_{nm}\frac{r^b_{m0}}{\omega_{m0}}+\frac{r^b_{nm}}{\omega_{nm}}r^d_{m0})+r^d_{n0}(-r^c_{0m}\frac{r^b_{mn}}{\omega_{mn}}+\frac{r^b_{0m}}{\omega_{0m}}r^c_{mn})$.

Note that $C^{n0}_{cbd}=r^c_{0n}(r^d_{n0})_{;b}$ and $S^{n0}_{abcd}=r^a_{0n;c}r^b_{n0;d}$, the last two terms in the braket is $i\partial_a(C^{n0}_{cbd}-C^{0n}_{dbc})+2iK^{0n}_{dcab}$. We have the following identities for the anti-symmetric part of $\partial C$:
\begin{align}
\frac{1}{2}\Big[\partial_a(C^{n0}_{cbd}-C^{0n}_{dbc})-\partial_b(C^{n0}_{cad}-C^{0n}_{dac})\Big]=\frac{1}{2}(K^{n0}_{cdab}+S^{n0}_{cdab}-S^{n0}_{cdba})-\frac{1}{2}(K^{0n}_{dcab}+S^{0n}_{dcab}-S^{0n}_{dcba})=-K^{0n}_{dcab}+S^{n0}_{cd[a,b]}.
\end{align}

Then the last two terms are: $i\partial_a(C^{n0}_{cbd}-C^{0n}_{dbc})+2iK^{0n}_{dcab}=i\text{d}C+iS^{n0}_{cd[a,b]}+iK^{0n}_{dcab}.$ The injection current tensor is:
\begin{align}
&\sigma^{abcd}_{\rm inj}(\omega)=\frac{\pi e^4}{\hbar^3V}\sum_n\delta(\omega_{n0}- \omega)\Bigg[iK^{0n}_{dcab}+i\text{d}C+iS^{n0}_{cd[a,b]}-2\partial_{a}(\omega_{n0})\partial_{E_b}Q^{n0}_{cd}\nonumber\\
&+2i\frac{\partial_a\omega_{n0}}{\omega_{n0}}\big(C^{n0}_{cdb}-C^{0n}_{dcb}\big)+2i\frac{\partial_a\omega_{n0}}{\omega_{n0}}\big(\frac{\partial_c\omega_{n0}}{\omega_{n0}}Q^{0n}_{db}-\frac{\partial_d\omega_{n0}}{\omega_{n0}}Q^{0n}_{bc}\big)\Bigg],
\end{align} 
where we have restored the physical constants.

\section{Shift current response}
The shift current response originates from the interband term and the anomalous velocity term:
\begin{align}
j^a_{\text{shift}}=\sum\limits_{bcd,nm}-E^br^b_{mn;a}\rho^{(2)}_{cd,nm}+\sum\limits_n-E^c\Omega^{ac}_{n0}(\rho^{(2)}_{bd,nn}+\rho^{(2)}_{db,nn})-E^d\Omega^{ad}_{n0}(\rho^{(2)}_{bc,nn}+\rho^{(2)}_{cb,nn}).
\end{align}

This also includes three parts:
\begin{enumerate}
    \item Contributions of $\nabla \rho^{(1)}_{n0/0n}$ to $\rho^{(2)}_{n0/0n}$ within the interband term.
    \item Contributions of commutator $[r,\rho^{(1)}]_{nm}$ to $\rho^{(2)}_{nm}$ within the interband term.
    \item Anomalous velocity term.
\end{enumerate}

Below we use the shorthand $\mathcal{C}_0=E^bE^cE^d$.

\subsection{$\nabla\rho^{(1)}$ contribution in the interband term}
Terms in this category are:
\begin{align}
&-i\sum\limits_{bcd,n}(E^br^b_{0n;a}\frac{E^c(\rho^{(1)}_{d,n0})_{;c}}{\omega_c+\omega_d-\omega_{n0}}+(0\leftrightarrow n))\nonumber\\
  &=-i\mathcal{C}_0\sum\limits_{bcd,n}\Big[\frac{r^b_{0n;a}}{\omega_c+\omega_d-\omega_{n0}}(\frac{r^d_{n0}}{
  \omega_d-\omega_{n0}
  })_{;c}-\frac{r^d_{n0;a}}{\omega_b+\omega_c-\omega_{0n}}(\frac{r^b_{0n}}{\omega_b-\omega_{0n}})_{;c}\Big].
\end{align}
We have resonant contributions from $\frac{1}{\omega_d-\omega_{n0}}$ or $\frac{1}{\omega_c-\omega_{0n}}$, from which we collect (the case with multiple poles will be treated separately):
\begin{align}
\pi\mathcal{C}_0\sum\limits_n\delta(\omega-\omega_{n0})\big[r^c_{0n;a}(\frac{r^b_{n0}}{\omega_{n0}})_{;d}+(\frac{r^b_{0n;a}}{\omega_{0n}})_{;c}r^d_{n0}-r^d_{n0;a}(\frac{r^b_{0n}}{\omega_{0n}})_{;c}-r^c_{0n}(\frac{r^b_{n0;a}}{\omega_{n0}})_{;d}\big]
\end{align}

Terms having double poles are (when the covariant derivative is on $b$):
\begin{align}
&-i\mathcal{C}_0\sum\limits_{n}\Big[\frac{r^c_{0n;a}}{\omega_d-\omega_{n0}}(\frac{r^d_{n0}}{
  \omega_d-\omega_{n0}
  })_{;b}-\frac{r^d_{n0;a}}{\omega_c-\omega_{0n}}(\frac{r^c_{0n}}{\omega_c-\omega_{0n}})_{;b}\Big]\nonumber\\
&=-i\mathcal{C}_0\sum\limits_n\Big[r^c_{0n;a}r^d_{n0;b}\frac{1}{(\omega_d-\omega_{n0})^2}-r^d_{n0;a}r^c_{0n;b}\frac{1}{(\omega_c-\omega_{0n})^2}+r^c_{0n;a}r^d_{n0}[\frac{1}{2(\omega_d-\omega_{n0})^2}]_{;b}-r^d_{n0;a}r^c_{0n}[\frac{1}{2(\omega_c-\omega_{0n})^2}]_{;b}\Big]
\end{align}

Note that $\frac{1}{(\omega-\omega_{n0}+i\eta)^2}=-\partial_{\omega}\frac{1}{\omega-\omega_{n0}+i\eta}=i\pi\partial_{\omega}\delta(\omega-\omega_{n0})+\text{non res.}$, and $\frac{1}{(-\omega-\omega_{0n}+i\eta)^2}=\partial_{\omega}\frac{1}{-\omega-\omega_{0n}+i\eta}=-{i\pi}\partial_{\omega}\delta(\omega-\omega_{n0})+\text{non res.}$, the resonant parts from the above can be written as:
\begin{align}
&{\pi C_0}\sum_n\partial_{\omega}\delta(\omega-\omega_{n0})\big[r^c_{0n;a}r^d_{n0;b}+r^d_{n0;a}r^c_{0n;b}-\frac{1}{2}(r^c_{0n;a}r^d_{n0})_{;b}-\frac{1}{2}(r^d_{n0;a}r^c_{0n})_{;b}\big].
\end{align}

In summary, we have:
\begin{align}
&j^a_{\text{shift}-1}=\pi\mathcal{C}_0\sum\limits_n\delta(\omega-\omega_{n0})\big[r^c_{0n;a}(\frac{r^b_{n0}}{\omega_{n0}})_{;d}+(\frac{r^b_{0n;a}}{\omega_{0n}})_{;c}r^d_{n0}-r^d_{n0;a}(\frac{r^b_{0n}}{\omega_{0n}})_{;c}-r^c_{0n}(\frac{r^b_{n0;a}}{\omega_{n0}})_{;d}\big]\nonumber\\
 &+{\pi C_0}\sum_n\partial_{\omega}\delta(\omega-\omega_{n0})\big[r^c_{0n;a}r^d_{n0;b}+r^d_{n0;a}r^c_{0n;b}-\frac{1}{2}(r^c_{0n;a}r^d_{n0})_{;b}-\frac{1}{2}(r^d_{n0;a}r^c_{0n})_{;b}\big]  
\end{align}

\subsection{Commutator contribution in the interband term}
The commutator contribution $[r,\rho^{(1)}]$ in $\rho^{(2)}$ for $n,m\neq 0$ is:
\begin{align}
 &  \sum\limits_{\text{cyc. } b} \mathcal{C}_0\sum\limits_{n,m}r_{mn;a}^b\big[ \frac{r^c_{n0}r^d_{0m}}{(\omega_c+\omega_d-\omega_{nm})(\omega_d-\omega_{0m})} + \frac{r^d_{n0}r^c_{0m}}{(\omega_c+\omega_d-\omega_{nm})(\omega_d-\omega_{n0})}+(c\leftrightarrow d)\big]\nonumber\\
&=\sum\limits_{\text{cyc. }b}\mathcal{C}_0\sum\limits_{n, m}r^b_{mn;a}\frac{1}{\omega_c+\omega_d-\omega_{nm}}\big[r^c_{n0}r^d_{0m}(\frac{1}{\omega_d-\omega_{0m}}+\frac{1}{\omega_c-\omega_{n0}})+r^d_{n0}r^c_{0m}(\frac{1}{\omega_d-\omega_{n0}}+\frac{1}{\omega_c-\omega_{0m}})]\nonumber\\
&=\sum\limits_{bcd}\mathcal{C}_0\sum\limits_{n, m}r^b_{mn;a}\frac{r^c_{n0}r^d_{0m}}{(\omega_c-\omega_{n0})(\omega_d-\omega_{0m})}
\end{align}

From which only the combinations $bdc,cdb,dbc$ have resonant contributions:
\begin{align}\label{eq:dipole_comm_p1}
&-i\pi\mathcal{C}_0\sum\limits_{n,m}r^b_{mn;a}r^d_{n0}r^c_{0m}\delta(\omega-\omega_{n0})\frac{1}{\omega_{mn}}-i\pi\mathcal{C}_0\sum\limits_{n,m}r^b_{mn;a}r^d_{n0}r^c_{0m}\delta(\omega-\omega_{m0})\frac{1}{\omega_{mn}}\nonumber\\
&+i\pi\mathcal{C}_0\sum\limits_{n,m}r^c_{mn;a}r^d_{n0}r^b_{0m}\delta(\omega-\omega_{n0})\frac{1}{\omega_{0m}}+i\pi\mathcal{C}_0\sum\limits_{n,m}r^d_{mn;a}r^b_{n0}r^c_{0m}\delta(\omega-\omega_{m0})\frac{1}{\omega_{n0}}
\end{align}

There are two extra contributions coming from $m=0$ and $n=0$:
\begin{align}\label{eq:dipole_comm_p2}
&\sum\limits_{bcd}\sum\limits_n-E^b(r^b_{n0;a}\rho^{(2)}_{cd,0n}+r^b_{0n;a}\rho^{(2)}_{cd,n0})=-\mathcal{C}_0\sum\limits_{bcd}\sum\limits_{m,n}\Big[\frac{r^b_{n0;a}r^d_{0m}r^c_{mn}}{(\omega_c+\omega_d-\omega_{0n})(\omega_d-\omega_{0m})}+\frac{r^b_{0n;a}r^c_{nm}r^d_{m0}}{(\omega_c+\omega_d-\omega_{n0})(\omega_d-\omega_{m0})}\Big].
\end{align}

From the first term on the RHS of Eq.~\eqref{eq:dipole_comm_p2}, only $bcd,dbc,dcb$ have resonant contributions:
\begin{align}\label{eq:dipole_comm_p2a}
&i\pi\mathcal{C}_0\sum\limits_{m,n}r^b_{n0;a}r^c_{0m}r^d_{mn}\delta(\omega-\omega_{m0})\frac{1}{\omega_{n0}}-i\pi\mathcal{C}_0\sum\limits_{m,n}r^d_{n0;a}r^b_{0m}r^c_{mn}\delta(\omega-\omega_{n0})\frac{1}{\omega_{0m}}\nonumber\\
&+i\pi\mathcal{C}_0\sum\limits_{m,n}r^d_{n0;a}r^c_{0m}r^b_{mn}\delta(\omega-\omega_{n0})\frac{1}{\omega_{mn}}-i\pi\mathcal{C}_0\sum\limits_{m,n}r^d_{n0;a}r^c_{0m}r^b_{mn}\delta(\omega-\omega_{m0})\frac{1}{\omega_{mn}}.
\end{align}

From the second term on the RHS of Eq.~\eqref{eq:dipole_comm_p2}, only $bcd,cbd,cdb$ have resonant contributions:
\begin{align}\label{eq:dipole_comm_p2b}
&i\pi\mathcal{C}_0\sum\limits_{m,n}r^b_{0n;a}r^c_{nm}r^d_{m0}\delta(\omega-\omega_{m0})\frac{1}{\omega_{0n}}+i\pi\mathcal{C}_0\sum\limits_{n,m}r^c_{0n;a}r^b_{nm}r^d_{m0}\delta(\omega-\omega_{n0})\frac{1}{\omega_{nm}}\nonumber\\
&-i\pi\mathcal{C}_0\sum\limits_{n,m}r^c_{0n;a}r^b_{nm}r^d_{m0}\delta(\omega-\omega_{m0})\frac{1}{\omega_{nm}}-i\pi\mathcal{C}_0\sum\limits_{n,m}r^c_{0n;a}r^d_{nm}r^b_{m0}\delta(\omega-\omega_{n0})\frac{1}{\omega_{m0}}.
\end{align}

Summing up Eq.~\eqref{eq:dipole_comm_p1}, Eq.~\eqref{eq:dipole_comm_p2a} and Eq.~\eqref{eq:dipole_comm_p2b}, we have:
\begin{align}
&j^a_{\text{shift}-2}=i\pi\mathcal{C}_0\sum\limits_n\delta(\omega-\omega_{n0})\Big\{r^d_{n0}\big[\frac{(r^b_{0m}r^c_{mn})_{;a}}{\omega_{0m}}-\frac{(r^c_{0m}r^b_{mn})_{;a}}{\omega_{mn}}\big]+r^c_{0n}\big[\frac{(r^d_{nm}r^b_{m0})_{;a}}{\omega_{m0}}-\frac{(r^b_{nm}r^d_{m0})_{;a}}{\omega_{nm}}\big]\nonumber\\
&+r^d_{n0;a}\big(r^c_{0m}\frac{r^b_{mn}}{\omega_{mn}}-\frac{r^b_{0m}}{\omega_{0m}}r^c_{mn}\big)+r^c_{0n;a}\big(\frac{r^b_{nm}}{\omega_{nm}}r^d_{m0}-r^d_{nm}\frac{r^b_{m0}}{\omega_{m0}}\big)  \Big\}.
\end{align}

\subsection{Anomalous velocity term}
We look at terms originated from: $j^a_{\text{shift}-3}=-E^c\sum\limits_n\Omega_{n0}^{ac}(\rho_{bd,nn}^{(2)}+\rho^{(2)}_{db,nn})-E^d\sum\limits_n\Omega_{n0}^{ad}(\rho_{bc,nn}^{(2)}+\rho^{(2)}_{cb,nn})$. Terms that have resonant contributions only come from $\rho^{(2)}_{bd,nn}$ and $\rho^{(2)}_{bc,nn}$:
\begin{align}
j^a_{\text{shift}-3}=E^bE^c\sum\limits_n\Omega_{n0}^{ac}(\rho^{(1)}_{d,n0}r^b_{0n})\frac{1}{\omega_b+\omega_d}-E^bE^d\sum\limits_n\Omega_{n0}^{ad}(r^b_{n0}\rho_{c,0n}^{(1)})\frac{1}{\omega_b+\omega_c}
\end{align}

The final answer is:
\begin{align}
    j^a_{\text{shift}-3}=-i\pi\frac{\mathcal{C}_0}{\omega_{n0}}\sum\limits_{n}\delta(\omega_{n0}-\omega)\big(\Omega^{ac}_{n0}r^d_{n0}r^b_{0n}-\Omega^{ad}_{n0}r^b_{n0}r^c_{0n}\big)
\end{align}

\subsection{Summary of third-order shift current}
Finally the shift current response is:
\begin{align}
&\sigma^{abcd}_{\rm shift}=\pi\sum\limits_n\delta(\omega-\omega_{n0})\Bigg\{i\sum\limits_{m}\Big[ r^d_{n0}\big(\frac{(r^{b}_{0m}r^c_{mn})_{;a}}{\omega_{0m}}-\frac{(r^c_{0m}r^b_{mn})_{;a}}{\omega_{mn}}\big)+ r^c_{0n}\big(\frac{(r^d_{nm}r^b_{m0})_{;a}}{\omega_{m0}}-\frac{(r^b_{nm}r^d_{m0})_{;a}}{\omega_{nm}}\big)\nonumber\\
&+ r^d_{n0;a}\big(r^c_{0m}\frac{r^b_{mn}}{\omega_{mn}}-\frac{r^b_{0m}}{\omega_{0m}}r^c_{mn}\big)+r^c_{0n;a}\big(\frac{r^b_{nm}}{\omega_{nm}}r^d_{m0}-r^d_{nm}\frac{r^b_{m0}}{\omega_{m0}}\big)\Big]\nonumber\\
&-i\frac{1}{\omega_{n0}}(\Omega^{ac}_{n0}r^d_{n0}r^b_{0n}-\Omega^{ad}_{n0}r^b_{n0}r^c_{0n})\nonumber\\
&+\bigg[r^c_{0n;a}\big(\frac{r^b_{n0}}{\omega_{n0}}\big)_{;d}+\big(\frac{r^b_{0n;a}}{\omega_{0n}}\big)_{;c}r^d_{n0}-r^d_{n0;a}\big(\frac{r^b_{0n}}{\omega_{0n}}\big)_{;c}-\big(\frac{r^b_{n0;a}}{\omega_{n0}}\big)_{;d}r^c_{0n}\bigg]
\Bigg\}\nonumber\\
&+\pi\sum\limits_n\partial_{\omega}\delta(\omega-\omega_{n0})\big[r^c_{0n;a}r^d_{n0;b}+r^d_{n0;a}r^c_{0n;b}-\frac{1}{2}(r_{0n;a}^cr^d_{n0})_{;b}-\frac{1}{2}(r^d_{n0;a}r^c_{0n})_{;b}\big].
\end{align}

Let's massage the second, third, fourth term into a form fully determined by resonant quantum geometry.

The third term can be simplified as:
\begin{align}
&r^c_{0n;a}\big(\frac{r^b_{n0}}{\omega_{n0}}\big)_{;d}+\big(\frac{r^b_{0n;a}}{\omega_{0n}}\big)_{;c}r^d_{n0}-r^d_{n0;a}\big(\frac{r^b_{0n}}{\omega_{0n}}\big)_{;c}-\big(\frac{r^b_{n0;a}}{\omega_{n0}}\big)_{;d}r^c_{0n}=\frac{1}{\omega_{n0}}(r^c_{0n;a}r^b_{n0;d}+r^d_{n0;a}r^b_{0n;c}-r^b_{0n;ac}r^d_{n0}-r^b_{n0;ad}r^c_{0n})\nonumber\\
&+\frac{\partial_c\omega_{n0}}{(\omega_{n0})^2}(r^b_{0n;a}r^d_{n0}-r^d_{n0;a}r^b_{0n})+\frac{\partial_d\omega_{n0}}{(\omega_{n0})^2}(-r^c_{0n;a}r^b_{n0}+r^b_{n0;a}r^c_{0n})
\end{align}

Adding up the second and the third term, we have:
\begin{align}
&\pi\sum\limits_n\delta(\omega_{n0}-\omega)\Big\{ \frac{1}{\omega_{n0}}\big[S^{n0}_{cb\{ad\}}+S^{n0}_{bd\{ac\}}-\partial_c(C^{0n}_{dab})-\partial_d(C^{n0}_{cab})-\frac{1}{2}(K^{0n}_{dbac}-K^{0n}_{bcad})\big]\\
&+\frac{1}{\omega_{n0}^2}\big[\partial_c\omega_{n0}(C^{0n}_{dab}-C^{n0}_{bad})+\partial_d\omega_{n0}(C^{n0}_{cab}-C^{0n}_{bac})\big]   \Big\}
\end{align}

The fourth term is:
\begin{align}
&\pi\sum\limits_n\partial_{\omega}\delta(\omega-\omega_{n0})\big[S^{n0}_{cd\{ab\}}-\frac{1}{2}\partial_b(C^{0n}_{dac}+C^{n0}_{cad})\big]
\end{align}

In summary, we have:
\begin{align}
&\sigma^{abcd}_{\rm shift}=\frac{\pi e^4}{\hbar^3 V}\sum\limits_n\delta(\omega-\omega_{n0})\Bigg\{i\sum\limits_{m}\Big[ r^d_{n0}\big(\frac{(r^{b}_{0m}r^c_{mn})_{;a}}{\omega_{0m}}-\frac{(r^c_{0m}r^b_{mn})_{;a}}{\omega_{mn}}\big)+ r^c_{0n}\big(\frac{(r^d_{nm}r^b_{m0})_{;a}}{\omega_{m0}}-\frac{(r^b_{nm}r^d_{m0})_{;a}}{\omega_{nm}}\big)\nonumber\\
&+ r^d_{n0;a}\big(r^c_{0m}\frac{r^b_{mn}}{\omega_{mn}}-\frac{r^b_{0m}}{\omega_{0m}}r^c_{mn}\big)+r^c_{0n;a}\big(\frac{r^b_{nm}}{\omega_{nm}}r^d_{m0}-r^d_{nm}\frac{r^b_{m0}}{\omega_{m0}}\big)\Big]\\
&+\frac{1}{\omega_{n0}^2}\big[\partial_c\omega_{n0}(C^{0n}_{dab}-C^{n0}_{bad})+\partial_d\omega_{n0}(C^{n0}_{cab}-C^{0n}_{bac})\big] \\
&+\frac{1}{\omega_{n0}}\big[S^{n0}_{cb\{ad\}}+S^{n0}_{bd\{ac\}}-\partial_c(C^{0n}_{dab})-\partial_d(C^{n0}_{cab})-\frac{1}{2}(K^{0n}_{dbac}-K^{0n}_{bcad})\big]\Bigg\}\\
&+\frac{\pi e^4}{\hbar^3 V}\sum\limits_n\partial_{\omega}\delta(\omega-\omega_{n0})\Big[S^{n0}_{cd\{ab\}}-\frac{1}{2}\partial_b(C^{0n}_{dac}+C^{n0}_{cad})\Big].
\end{align}

For excitons, several quantum geometric terms vanish identically because $K=0$, $\partial_a{\omega_{n0}}=0$ and $\partial_a(C^{n0}_{bcd})=0$. The excitonic shift current is therefore:
In summary, we have:
\begin{align}
&\sigma^{abcd}_{\rm shift,exciton}=\frac{\pi e^4}{\hbar^3 V}\sum\limits_n\delta(\omega-\omega_{n0})\Bigg\{i\sum\limits_{m}\Big[ r^d_{n0}\big(\frac{(r^{b}_{0m}r^c_{mn})_{;a}}{\omega_{0m}}-\frac{(r^c_{0m}r^b_{mn})_{;a}}{\omega_{mn}}\big)+ r^c_{0n}\big(\frac{(r^d_{nm}r^b_{m0})_{;a}}{\omega_{m0}}-\frac{(r^b_{nm}r^d_{m0})_{;a}}{\omega_{nm}}\big)\nonumber\\
&+ r^d_{n0;a}\big(r^c_{0m}\frac{r^b_{mn}}{\omega_{mn}}-\frac{r^b_{0m}}{\omega_{0m}}r^c_{mn}\big)+r^c_{0n;a}\big(\frac{r^b_{nm}}{\omega_{nm}}r^d_{m0}-r^d_{nm}\frac{r^b_{m0}}{\omega_{m0}}\big)\Big]\\
&+\frac{1}{\omega_{n0}}\big(2S^{n0}_{cbad}+2S^{n0}_{bdac}\big)\Bigg\}+\frac{2\pi e^4}{\hbar^3 V}\sum\limits_n\partial_{\omega}\delta(\omega-\omega_{n0})S^{n0}_{cdab}.
\end{align}

We can make one further simplification of the first term by partial derivatives:
\begin{align}
 &\sum\limits_{m}i r^d_{n0}\big(\frac{(r^{b}_{0m}r^c_{mn})_{;a}}{\omega_{0m}}-\frac{(r^c_{0m}r^b_{mn})_{;a}}{\omega_{mn}}\big)+\sum\limits_{m}i r^c_{0n}\big(\frac{(r^d_{nm}r^b_{m0})_{;a}}{\omega_{m0}}-\frac{(r^b_{nm}r^d_{m0})_{;a}}{\omega_{nm}}\big)\nonumber\\
&+\sum\limits_{m}i r^d_{n0;a}\big(r^c_{0m}\frac{r^b_{mn}}{\omega_{mn}}-\frac{r^b_{0m}}{\omega_{0m}}r^c_{mn}\big)+\sum\limits_{m}i r^c_{0n;a}\big(\frac{r^b_{nm}}{\omega_{nm}}r^d_{m0}-r^d_{nm}\frac{r^b_{m0}}{\omega_{m0}}\big)\nonumber\\
&=i\sum\limits_m\big[r^d_{n0}(\frac{r^b_{0m}r^c_{mn}}{\omega_{0m}}-\frac{r^c_{0m}r^b_{mn}}{\omega_{mn}})+r^c_{0n}(\frac{r^d_{nm}r^b_{m0}}{\omega_{m0}}-\frac{r^b_{nm}r^d_{m0}}{\omega_{nm}})\big]_{;a}\nonumber\\
&+i\sum\limits_m[r^d_{n0}(\frac{r^b_{0m}r^c_{mn}}{\omega_{0m}^2}v^a_{0m}-\frac{r^c_{0m}r^b_{mn}}{\omega_{mn}^2}v^a_{mn})+r^c_{0n}(\frac{r^d_{nm}r^b_{m0}}{\omega_{m0}^2}v^a_{m0}-\frac{r^b_{nm}r^d_{m0}}{\omega_{nm}^2}v^a_{nm})]\nonumber \\
&+2i\sum\limits_{m} [r^d_{n0;a}\big(r^c_{0m}\frac{r^b_{mn}}{\omega_{mn}}-\frac{r^b_{0m}}{\omega_{0m}}r^c_{mn}\big)+ r^c_{0n;a}\big(\frac{r^b_{nm}}{\omega_{nm}}r^d_{m0}-r^d_{nm}\frac{r^b_{m0}}{\omega_{m0}}\big)].
\end{align}
The first term vanishes because the quantity inside the generalized derivative is gauge invariant and, after summing over all the intermediate states $m$, for a localized exciton, independent of the flux $\kappa$. 

The simplified exciton shift current expression is therefore (noticing that $(r^d_{n0})_{;a}=-i\mathcal{R}^a_{n0}r^d_{n0}$):

\begin{align}
&\sigma^{abcd}_{\rm shift,exciton}=\frac{\pi e^4}{\hbar^3 V}\sum\limits_n\delta(\omega-\omega_{n0})\Bigg\{i\sum\limits_mv^a_m\Big[r^c_{0n}(\frac{r^d_{nm}r^b_{m0}}{\omega_{m0}^2}+\frac{r^b_{nm}r^d_{m0}}{\omega_{nm}^2})-r^d_{n0}(\frac{r^b_{0m}r^c_{mn}}{\omega_{0m}^2}+\frac{r^c_{0m}r^b_{mn}}{\omega_{mn}^2})\Big]\nonumber\\
&-2\mathcal{R}^a_{n0} \partial_{E_b}Q^{n0}_{cd}+\frac{2}{\omega_{n0}}\big(S^{n0}_{cbad}+S^{n0}_{bdac}\big)\Bigg\}+\frac{2\pi e^4}{\hbar^3 V}\sum\limits_n\partial_{\omega}\delta(\omega-\omega_{n0})S^{n0}_{cdab}.
\end{align}

\section{Proof that non-resonant terms vanish}
In the above sections we analyzed on-resonant terms (injection and shift) to the third order photoconductivity, now we examine the principal parts in Eq.~\ref{eq:diag_derivative},~\ref{eq:diag_comm},~\ref{eq:third_order_current_inter},~\ref{eq:third_order_current_anom}. As we will now show, in the limit of $\omega_{\Sigma}\rightarrow 0$, the third order conductivity arising from such off-resonant terms vanishes. To see this, we show $\sigma^{(3)}/\omega_{\Sigma}$ is finite when off-resonant. As a result, $\sigma^{(3)}\rightarrow 0$ when $\omega_{\Sigma}\rightarrow 0$. Below we explicitly preserve $\omega_b,\omega_c,\omega_d$ and use $\omega_{\Sigma}=\omega_b+\omega_c+\omega_d$.

The full expression is composed of five terms. The first term is the pure commutator contribution in the diagonal-commutator.
\begin{align}
   & - \mathcal{C}_0\frac{1}{\omega_{\Sigma}}\sum\limits_{n,m}\partial_a(\omega_{m0})\frac{r^b_{nm}r^c_{m0}r^d_{0n}}{(\omega_d-\omega_{0n})(\omega_b+\omega_d-\omega_{0m})(\omega_c-\omega_{m0})}+\mathcal{C}_0\frac{1}{\omega_{\Sigma}}\sum\limits_{n,m}\partial_a(\omega_{n0})\frac{r^b_{nm}r^c_{m0}r^d_{0n}}{(\omega_d-\omega_{0n})(\omega_c-\omega_{m0})(\omega_b+\omega_c-\omega_{n0})}
\end{align}

The second term is the diagonal-derivative, plus $\nabla\rho^{(1)}_{n0}$ in the diagonal-commutator:
\begin{align}
   & \frac{\mathcal{C}_0}{-i\omega_{\Sigma}^2}\sum\limits_{bcd}\partial_a(\omega_{n0})\Big[ \frac{r^b_{n0}}{\omega_c+\omega_d-\omega_{0n}}(\frac{r^d_{0n}}{\omega_d-\omega_{0n}})_{;c} +(\frac{r^b_{n0}}{\omega_b-\omega_{n0}})_{;c}\frac{r^d_{0n}}{\omega_c+\omega_b-\omega_{n0}}+\nabla_c\Big(\frac{r^b_{n0}r^d_{0n}}{(\omega_b-\omega_{n0})(\omega_d-\omega_{0n})}\Big)\Big]\nonumber\\
    &=\frac{\mathcal{C}_0}{-i\omega_{\Sigma}}\sum\limits_{bcd}\partial_a(\omega_{n0})\Big[(\frac{r^d_{0n}}{\omega_d-\omega_{0n}})_{;c}\frac{r^b_{n0}}{(\omega_c+\omega_d-\omega_{0n})(\omega_b-\omega_{n0})}+(\frac{r^b_{n0}}{\omega_b-\omega_{n0}})_{;c}\frac{r^d_{0n}}{(\omega_c+\omega_b-\omega_{n0})(\omega_d-\omega_{0n})}\Big]
\end{align}

The third term is the commutator contribution in the interband term, which is:
\begin{align}
& \frac{1}{\omega_{\Sigma}}\sum\limits_{bcd}\mathcal{C}_0\sum\limits_{n, m}r^b_{mn;a}\frac{r^c_{n0}r^d_{0m}}{(\omega_c-\omega_{n0})(\omega_d-\omega_{0m})}   -\frac{1}{\omega_{\Sigma}}\mathcal{C}_0\sum\limits_{bcd}\sum\limits_{m,n}\Big[\frac{r^b_{n0;a}r^d_{0m}r^c_{mn}}{(\omega_c+\omega_d-\omega_{0n})(\omega_d-\omega_{0m})}+\frac{r^b_{0n;a}r^c_{nm}r^d_{m0}}{(\omega_c+\omega_d-\omega_{n0})(\omega_d-\omega_{m0})}\Big]\nonumber\\
&=\frac{1}{\omega_{\Sigma}}\mathcal{C}_0\sum\limits_{bcd}\sum\limits_{n,m}\Big[r^b_{nm;a}\frac{r^c_{m0}r^d_{0n}}{(\omega_c-\omega_{m0})(\omega_d-\omega_{0n})}-\frac{r^c_{m0;a}r^d_{0n}r^b_{nm}}{(\omega_b+\omega_d-\omega_{0m})(\omega_d-\omega_{0n})}-\frac{r^d_{0n;a}r^b_{nm}r^c_{m0}}{(\omega_b+\omega_c-\omega_{n0})(\omega_c-\omega_{m0})}\Big].
\end{align}

The fourth term is the $\nabla \rho^{(1)}$ contribution in the interband term:
\begin{align}
\frac{-i}{\omega_{\Sigma}}\mathcal{C}_0\sum\limits_{bcd,n}\Big[\frac{r^d_{0n;a}}{\omega_c+\omega_b-\omega_{n0}}(\frac{r^b_{n0}}{
  \omega_b-\omega_{n0}
  })_{;c}-\frac{r^b_{n0;a}}{\omega_d+\omega_c-\omega_{0n}}(\frac{r^d_{0n}}{\omega_d-\omega_{0n}})_{;c}\Big]
\end{align}

The fifth term is the anomalous velocity term:
\begin{align}
\frac{\mathcal{C}_0}{\omega_{\Sigma}}\sum\limits_{bcd,n}\Omega^{ac}_{n0}\frac{r^b_{n0}r^d_{0n}}{(\omega_b-\omega_{n0})(\omega_d-\omega_{0n})}.
\end{align}

After some lengthy algebra, we can combine all these terms into the following expression:
\begin{align}
&\sum\limits_{bcd}\sum\limits_{n,m}-\mathcal{C}_0(\frac{r^c_{m0}}{\omega_c-\omega_{m0}})_{;a}\frac{r^b_{nm}r^d_{0n}}{(\omega_b+\omega_d-\omega_{0m})(\omega_d-\omega_{0n})}-\mathcal{C}_0(\frac{r^d_{0n}}{\omega_d-\omega_{0n}})_{;a}\frac{r^b_{nm}r^c_{m0}}{(\omega_c+\omega_b-\omega_{n0})(\omega_c-\omega_{m0})}\nonumber\\
&\sum\limits_{bcd}\sum\limits_{n}-i\mathcal{C}_0(\frac{r^d_{0n}}{\omega_d-\omega_{0n}})_{;a}(\frac{r^b_{n0}}{\omega_b-\omega_{n0}})_{;c}\frac{1}{\omega_b+\omega_c-\omega_{n0}}+i\mathcal{C}_0(\frac{r^d_{0n}}{\omega_d-\omega_{0n}})_{;c}(\frac{r^b_{n0}}{\omega_b-\omega_{n0}})_{;a}\frac{1}{\omega_c+\omega_d-\omega_{0n}}.
\end{align}
This is finite when off-resonant, therefore $\sigma^{(3)}\rightarrow 0$ when $\omega_{\Sigma}\rightarrow 0$ in the off-resonant regime.

\section{Numerical details}
We study a three-band tight-binding model for MoS$_2$ subject to varying strains. The hopping parameters and the spin-orbit coupling strength are adapted from Ref.~\cite{liu2013three}. We use a screened Coulomb potential of the Keldysh form $V_R=\frac{\pi e^2}{2\epsilon r_0}[H_0(R/r_0)-Y_0(R/r_0)]$ following Ref.~\cite{wu2015exciton}, with $r_0=33.875\,\mathring{\text{A}}/\epsilon$ and $\epsilon=2.5$. The strain is applied along armchair direction ($y$ direction in the coordinate system we chose) and is incorporated by modifying the hopping strength via $t(\textbf{r})=t(\textbf{r}_0)(1-\Lambda \frac{|\textbf{r}-\textbf{r}_0|}{|\textbf{r}_0|})$ with $\Lambda=5$, following Ref.~\cite{rostami2015theory}.

Throughout our calculation, the following parameters are used. The lattice constant is $a=3.16\,\mathring{\text{A}}$. The static electric field is $10^5$ V/m, the electric field of light is $3\times 10^5$ V/m and the relaxation time is $\tau=100$ fs for injection current calculation. We use the regularized form of $\delta$ function: $\delta(\hbar\omega-E_{n0})=\frac{\Gamma}{\pi}\frac{1}{(\hbar\omega-E_{n0})^2+\Gamma^2}$, and $\frac{1}{\hbar}\partial_{\omega}\delta(\hbar\omega-E_{n0})=-\frac{2\Gamma}{\pi}\frac{\hbar\omega-E_{n0}}{[(\hbar\omega-E_{n0})^2+\Gamma^2]^2}$, where $\Gamma=25$ meV.

For injection current, we study the Hall current under a circularly polarized light. This corresponds to the tensor $\sigma^C_{\text{Hall}}=i(\sigma^{xyxy}-\sigma^{yxxy}-\sigma^{xyyx}+\sigma^{yxyx})/2$. In the non-interacting case, we have chosen photon energy to be $\hbar\omega=2.6\,\text{eV}$ and the response converges on a $300\times 300 $ lattice. For exciton, it has been established in Ref.~\cite{yang2026correlated} that $\partial_{\bm{\kappa}}E_{n0}=0$, and $\partial_aQ^{n_{\rm ex}0}_{bc}=0$ at $\bm{\kappa}=0$ hold in the thermodynamic limit, therefore we make use of it to simplify the injection current expression. The resulting expression only involves $K$ and is of the form: $\sigma^C_{\text{Hall}}=\frac{\pi e^4}{\hbar^3V}\delta(\omega-\omega_{n0})(K^{n0}_{xyxy}-K^{n0}_{yxxy})$. In the exciton case, we have chosen the photon energy at the lowest exciton energy. The response is numerically zero on a $60\times 60$ lattice.

For shift current, we study the longitudinal current under a linearly polarized light. In the non-interacting case, we have chosen photon energy to be $\hbar\omega=2.6\,\text{eV}$ and the response converges on a $750\times 750$ lattice. In the exciton case, we choose the photon energy at $10\,\text{meV}$ above the lowest exciton energy to ensure an appreciable contribution from $\partial_\omega\delta(\omega-\omega_{n0})$ term. The response converges on a $78\times 78$ lattice. At zero strain for exciton, Ref.~\cite{yang2026correlated} showed that the excitonic transition shift vector vanishes by $C_{3z}$ symmetry. Since the excitonic Hermitian connection is proportional to the excitonic transition shift vector, it vanishes as well. Using the identity $Q_{ab}^{n0} S_{abcd}^{n0}=(C_{bca}^{n0})^*C_{adb}^{n0}$, it then follows that $S$ also vanishes. Finally, the velocity of each eigenstate transforms as a vector and must vanish under $C_{3z}$, thereby eliminating the term $\mathcal{V}$ in Eq.~\eqref{eq:shift_exciton_formula}. Consequently, the third-order excitonic shift current vanishes at zero strain, consistent with our numerical results.

\end{widetext}
\end{document}